\documentclass[journal]{IEEEtran}
\usepackage{cite}
\usepackage{amsmath,amssymb,amsfonts}
\usepackage{algorithm, algorithmicx, algpseudocode}
\usepackage{graphicx}
\usepackage{textcomp}
\usepackage{xcolor}
\usepackage{bm}
\usepackage{makecell}
\usepackage{balance}
\usepackage{booktabs}
\usepackage{siunitx}
\usepackage{stfloats}
\usepackage{bbding}
\usepackage{url}
\usepackage[caption=false,font=footnotesize,labelfont=rm,textfont=rm]{subfig}

\usepackage[colorlinks,
            linkcolor=blue,       
            anchorcolor=blue,  
            citecolor=blue,        
            ]{hyperref}
\def\BibTeX{{\rm B\kern-.05em{\sc i\kern-.025em b}\kern-.08em
    T\kern-.1667em\lower.7ex\hbox{E}\kern-.125emX}}
\usepackage{lineno} 
\begin{document}
\title{CSI Transfer From Sub-6G to mmWave: Reduced-Overhead Multi-User Hybrid Beamforming}
\author{
    Weicao Deng,~\IEEEmembership{Student Member,~IEEE}, 
    Min Li,~\IEEEmembership{Member,~IEEE}, 
    Ming-Min Zhao,~\IEEEmembership{Senior Member,~IEEE},\\
    Min-Jian Zhao,~\IEEEmembership{Member,~IEEE},
    and Osvaldo Simeone,~\IEEEmembership{Fellow,~IEEE}
\thanks{
Weicao Deng, Min Li, Ming-Min Zhao, and Min-Jian Zhao are with the College of Information Science and Electronic Engineering and also with Zhejiang Provincial Key Laboratory of
Information Processing, Communication and Networking (IPCAN), Zhejiang University, Hangzhou 310027, China (e-mail: \{caowd,~min.li,~zmmblack,~mjzhao\}@zju.edu.cn). (\emph{Corresponding authors: Min Li; Ming-Min Zhao.})

\par Osvaldo Simeone is with the Department of Engineering, King’s College London, London WC2R 2LS, UK (e-mail: osvaldo.simeone@kcl.ac.uk).
\par 
The work of Osvaldo Simeone has been supported by an Open Fellowship of the EPSRC with reference EP/W024101/1, by the EPSRC project EP/X011852/1, by the European Union’s Horizon Europe Project CENTRIC under Grant 101096379. 
}}

\maketitle
\begin{abstract}
Hybrid beamforming is vital in modern wireless systems, especially for massive MIMO and millimeter-wave (mmWave) deployments, offering efficient directional transmission with reduced hardware complexity.
However, effective beamforming in multi-user scenarios relies heavily on accurate channel state information, the acquisition of which often requires significant pilot overhead, degrading system performance. 
To address this and inspired by the spatial congruence between sub-6GHz (sub-6G) and mmWave channels, we propose a Sub-6G information Aided Multi-User Hybrid Beamforming (SA-MUHBF) framework, avoiding excessive use of pilots at mmWave. 
SA-MUHBF employs a convolutional neural network to predict mmWave beamspace from sub-6G channel estimate, followed by a novel multi-layer graph neural network for analog beam selection and a linear minimum mean-square error algorithm for digital beamforming.
Numerical results demonstrate that SA-MUHBF efficiently predicts the mmWave beamspace representation and achieves superior spectrum efficiency over state-of-the-art benchmarks. Moreover, SA-MUHBF demonstrates robust performance across varied sub-6G system configurations and exhibits strong generalization to unseen scenarios.

\end{abstract}

\begin{IEEEkeywords}
Millimeter-wave communication, hybrid beamforming, sub-6G channel, deep learning, graph neural network.
\end{IEEEkeywords}

\section{Introduction}
Millimeter-wave (mmWave) communication has emerged as a promising technology for high-speed wireless systems, particularly in multi-user environments. 
With its abundant bandwidth, mmWave enables high data rates to meet the increasing demand for multi-user applications \cite{8454665}.
However, mmWave communication faces significant challenges such as severe path loss and blockages caused by obstacles.
To address these challenges and improve system performance, the massive multiple input multiple output (MIMO) technology is widely employed in mmWave communication systems due to its ability to provide large array gain, and beamforming techniques have thus become increasingly important.

\subsection{Related Works and Motivations}
Among various beamforming architectures, the hybrid beamforming architecture garners considerable attention owing to its balanced performance in complexity and power consumption \cite{Alkhateeb2014,heath2016overview}.
The authors in \cite{sohrabiHybridDigitalAnalog2016} demonstrated that if the number of RF chains is twice the total number of data streams, the hybrid beamforming structure can realize any fully digital beamformer exactly.
A significant number of works \cite{sohrabiHybridDigitalAnalog2016,alkhateeb2016frequency,huTwoTimescaleEndtoEndLearning2022} investigated hybrid beamforming in single user scenarios and proposed a range of solutions based on both optimization \cite{sohrabiHybridDigitalAnalog2016,alkhateeb2016frequency} and deep learning \cite{huTwoTimescaleEndtoEndLearning2022}.
Moreover, in multi-user scenario, reference \cite{alkhateeb2015limited} developed a two-stage hybrid beamforming algorithm by separating analog and digital beamforming design based on limited feedback of channel state information (CSI), while reference \cite{9045972} extended this two-stage method into a beam pairing algorithm to reduce inter-beam interference by exploring partial interfering beam feedback.
Reference \cite{10187715} proposed an efficient iterative algorithm for sum-utility maximization by utilizing the alternating minimization and manifold optimization methods.
Reference \cite{elbir2019hybrid} employed a deep neural network to learn the optimal analog beam selection labels generated through exhaustive search.
Reference \cite{jinModelDrivenDeepLearning2023} investigated the hybrid beamforming problem in a multi-user MIMO system and proposed a model-driven deep learning algorithm.
Nevertheless, it is noted that the aforementioned methods only leverage channel measurements in the mmWave band and may fail to achieve good performance with limited pilot budget or with low signal-to-noise ratio (SNR), particularly in multi-user scenarios.

\par More recently, researchers have explored the potential of using sub-6 GHz (sub-6G) channel data to enhance mmWave communication, supported by multiple channel measurement campaigns \cite{peter2016measurement,samimi20163,9768944,10058899}.
In particular, the indoor channel measurements in \cite{peter2016measurement} showed that the angle power profiles of 5.8GHz, 14.8GHz, and 58.7GHz channels share the same scatters that result in similar spatial information, and the similarity was furthur verified in an outdoor scenario reported in \cite{samimi20163}.
Reference \cite{9768944} presented an extensive simultaneous multi-band measurement campaign in an industrial hall covering the sub-6 GHz, 30 GHz and 60 GHz bands, and the results demonstrated that sub-6G and mmWave channels share similar geometric properties.
Furthermore, the point cloud ray-tracing and propagation measurement results on 4, 15, 28, 60, and 86 GHz in \cite{10058899} also demonstrated the feasibility of ultilizing low-frequency radio channel information for high-frequency beam search.

\par Motivated by the aforementioned spatial congruence characteristic, an increasing amount of works have been dedicated to enhancing mmWave communication with sub-6G channel information.
Specifically, reference \cite{nitsche2015steering} proposed to utilize the 2.4/5 GHz channel information for directional mmWave link establishment and demonstrated its effectiveness by constructing a practical system.
Reference \cite{7888146} provided both non-parametric and parametric approaches to obtain mmWave channel covariance matrix from sub-6G channel information.
Reference \cite{8792393} further developed a covariance translation approach and presented an out-of-band (OOB) aided compressed covariance estimation scheme.
Reference \cite{ali2017millimeter} proposed a classical logit weighted orthogonal matching pursuit (LW-OMP) algorithm that utilizes sub-6G channel information to assist mmWave beam selection.
Additionally, a series of deep-learning techniques have been proposed to facilitate beam selection \cite{alrabeiah2020deep,10292615,10229493,dengwcnc2024} and beam tracking \cite{ma2021deep}.
Nevertheless, most of the above works only focused on the single user scenario and considered the fully analog beamforming architecture, thereby restricting their potential application in multi-user scenarios.

\par Different from the single user scenario where the spatial congruence between sub-6G and mmWave channels is primarily used to determine the dominant path direction, leveraging sub-6G channel information in mmWave multi-user scenarios necessitates a more comprehensive consideration of interference coordination and resource allocation among users.
The authors in \cite{maschiettiCoordinatedBeamSelection2019} investigated the utilization of sub-6G channel information for multi-user mmWave hybrid beamforming and developed both uncoordinated and coordinated methods to select analog beams based on sub-6G channel estimates.
Reference \cite{liHybridPrecodingUsing2020} extended the uncoordinated method by considering a more efficient Grassmannian training codebook.
Both studies in \cite{maschiettiCoordinatedBeamSelection2019} and \cite{liHybridPrecodingUsing2020} performed beam selection directly based on sub-6G channel estimates, but this approach might be affected by the variation between sub-6G and mmWave channels.
Reference \cite{liuCollaborativeManagementResource2023} investigated the resource allocation and precoding problem in the dual-mode network and proposed a rapid frequency band allocation and precoding algorithm that leverages the spatial similarity between sub-6G and mmWave channels.
However, the interference among UEs is not considered in \cite{liuCollaborativeManagementResource2023} as the UEs are assigned to different sub-frequency bands. 
It is further noted that the above works lack dedicated preprocessing of sub-6G information and perform only simple or no interference coordination among UEs.

\par Recently, graph neural networks (GNNs) have garnered significant attention in the wireless communication field due to their superior capability in processing non-Euclidean data, e.g., CSI.
In particular, a significant number of works have appied GNNs to solve resource allocation and interference coordination problems \cite{he2021overview,shen2020graph,lee2020graph,he2022gblinks,10184114}.
Reference \cite{shen2020graph} demonstrated that radio resource management problems can be formulated as graph optimization problems and developed a family of neural networks, named MPGNNs, to solve them.
Reference \cite{lee2020graph} investigated the link scheduling problem in device-to-device (D2D) networks and developed a graph embedding based method to extract the interference pattern for each D2D pair to perform link scheduling.
Reference \cite{he2022gblinks} further considered the joint beam selection and link activation problem in ultra-dense D2D mmWave communication networks and designed a deep learning architecture based on GNN, named GBLinks.
The more recent work \cite{10184114} studied the joint user association and beam selection problem for mmWave-integrated heterogeneous networks and developed a GNN-aided algorithm using a primal-dual learning framework.
In the aforementioned works, GNNs are constructed with multi-layer perceptrons (MLPs) or 1D convolutional neural networks (CNNs), which neglect the cluster characteristic of MIMO beamspace channels and therefore may degrade the feature extraction efficiency.
Besides, the above GNNs operate directly on the graph constructed based on the CSI data, lacking a dedicated pre-processing phase for the inputs to further enhance the overall performance.

\subsection{Main Contributions} 
Despite the previous research efforts, leveraging sub-6G channel information to enhance mmWave multi-user hybrid beamforming remains a significant challenge. 
This challenge stems from the need to efficiently transfer sub-6G CSI to mmWave frequencies and to manage inter-user interference effectively to achieve optimal performance. 
Although GNNs have been effective in managing interference across various applications, they have not yet been specifically adapted for the task of multi-user hybrid beamforming. 
In this paper, we address these challenges by proposing a Sub-6G information Aided Multi-User Hybrid Beamforming (SA-MUHBF) framework. 
Our approach uses CNNs to accurately predict the mmWave beamspace from sub-6G channel estimates and proposes a customized GNN for analog beam selection. 
The main contributions of this paper are summarized as follows:

\begin{enumerate}
    \item We propose a three-stage SA-MUHBF framework that consists of a mmWave beamspace representation prediction stage, an analog beam selection stage, and a digital beamforming stage. The proposed framework first predicts the mmWave beamspace representation from sub-6G channel estimates, and then applies a decoupled beamforming design, where the predicted mmWave beamspace representation is used for analog beam selection, and a linear minimum mean squared error (LMMSE) algorithm is adopted for digital beamforming design.
    \item We propose a 2D CNN architecture in the mmWave beamspace representation prediction stage, which is able to effectively leverage the cluster characteristic of MIMO beamspace channels, thereby facilitating the feature extraction from sub-6G channel information.
    \item Taking the inter-user interference coordination into account, we design a multi-layer GNN architecture to iteratively improve the quality of beam selection. In particular, we propose a novel graph convolution layer that consists of a preprocessing phase and a 2D CNN-based graph convolution. The preprocessing phase explicitly captures the effective signals and interference among links, and then the 2D CNNs in graph convolution effectively extract the messages from them and perform beam selection strategy update.
    \item Comprehensive numerical simulations are performed to validate the efficiency of the proposed SA-MUHBF framework, as compared to the existing contemporary benchmarks. The scalability of the SA-MUHBF is also verified under various sub-6G system configurations. Additionally, the generalization capabilities of the SA-MUHBF are also demonstrated with test datasets generated from previously unexplored regions.
\end{enumerate}

\par The remainder of this paper is organized as follows. 
In Section~\ref{system_model_problem_formulation}, we begin by introducing the system model, and then proceed to formulate the considered sub-6G aided mmWave multi-user hybrid beamforming problem.
In Section~\ref{Framework_of_Sub-6G_Information-Aided_Multi-User_Hybrid_Beamforming}, we discuss and summarize the key challenges to solve this problem and propose the overall design of SA-MUHBF framework.
Section~\ref{Architecture_and_Implementation_of_Sub-6G_Information-Aided_Multi-User_Hybrid_Beamforming} provides the detailed architecture and implementation of SA-MUHBF.
Following that, Section~\ref{Numerical_Results} presents the detailed numerical results that demonstrate the effectiveness of the proposed SA-MUHBF.
Finally, the conclusion is drawn in Section~\ref{Conclusion}. 

\subsubsection*{Notations} Scalars, vectors and matrices are respectively denoted by lower/upper case, boldface lower case and boldface upper case letters.
Notation ${\mathbf I}_m$ represents an $m\times m$ identify matrix.
Superscripts $(\cdot)^*$, $(\cdot)^T$, $(\cdot)^H$, $(\cdot)^{-1}$, and $(\cdot)^{\dagger}$ are used to denote the conjugate, transpose, conjugate transpose, inverse, and pseudo-inverse operations, respectively.
Operators $\odot$, ${\mathbb E}(\cdot)$, $|.|$, $\left\lVert.\right\lVert_{\ell_0}$, $\left\lVert.\right\lVert_{\ell_1}$ and $\left\lVert.\right\lVert_{\ell_2}$ represent the dot product, expectation, absolute value, $\ell_0$-norm, $\ell_1$-norm, and $\ell_2$-norm, respectively.
$\mathcal{CN}(0,\sigma^2)$ is a zero-mean complex Gaussian \textcolor{black}{distribution} with variance $\sigma^2$.
Moreover, to distinguish between the sub-6G system and mmWave system, we use $\underline{(\cdot)}$ to indicate parameters corresponding to the sub-6G system, as exemplified by $\underline{x}$.

\section{System Model and Problem Formulation}
\label{system_model_problem_formulation}
\begin{figure*}[t]
    \centering
    \centering
        \subfloat[]{  
                \includegraphics[width=0.48\textwidth]{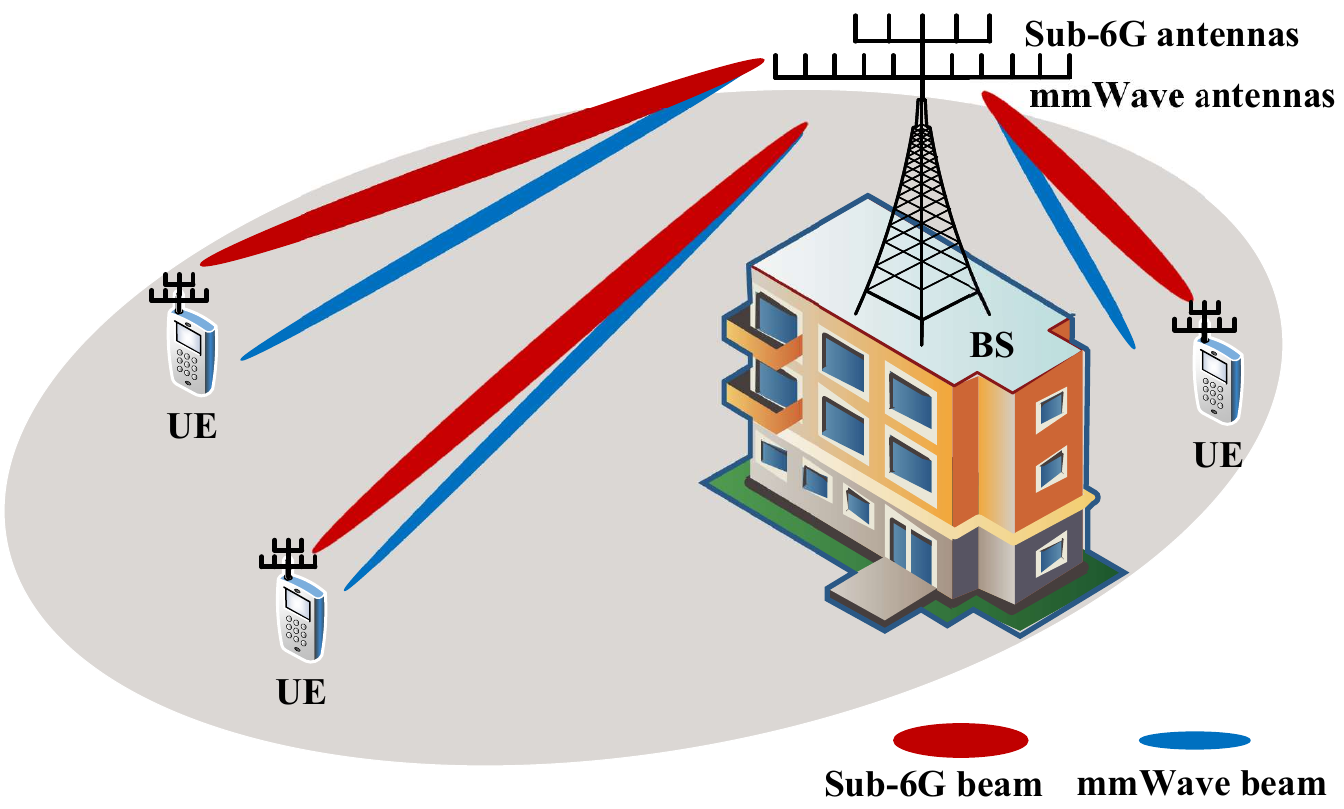}  
                \label{system model:sub1}  
            } 
            \subfloat[]{  
                \includegraphics[width=0.48\textwidth]{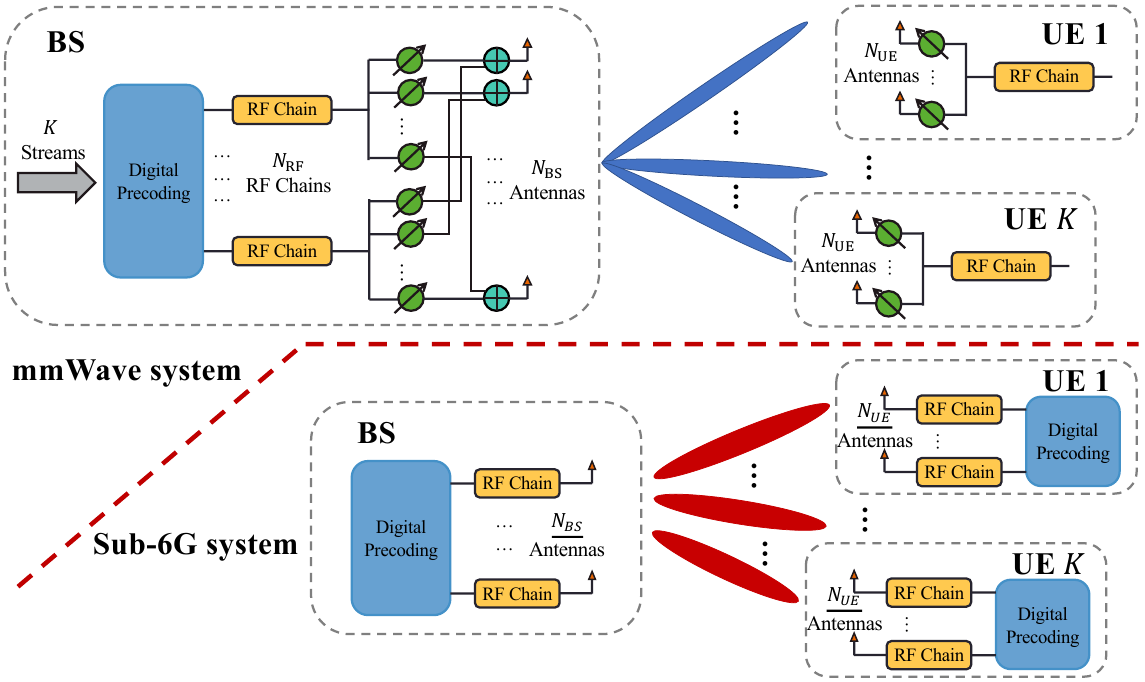}  
                \label{system model:sub2}  
            } 
        \caption{Illustration of (a) the dual-band communication system and (b) the beamforming architectures of mmWave system and sub-6G system.}
        \label{system model}
    \end{figure*}
As depicted in Fig.~\ref{system model:sub1}, we consider a dual-band system comprising one BS and $K$ UEs.
Both the BS and UEs are assumed to employ two transceivers similar to \cite{ali2017millimeter,alrabeiah2020deep,liuCollaborativeManagementResource2023}: one works at sub-6G frequency, and the other operates at mmWave frequency.
Specifically, in the mmWave system, the BS adopts a hybrid beamforming architecture, where each of the $N_{\text{BS}}$ mmWave antennas is connected to $N_{\text{RF}}\geqslant K$ RF chains.
Each UE adopts an analog beamforming architecture, where $N_{\text{UE}}$ mmWave antennas are fully connected with one RF chain.
As for the sub-6G system, both the BS and UEs adopt the fully digital beamforming architecture and are equipped with $\underline{N_{\text{BS}}}$ and $\underline{N_{\text{UE}}}$ antennas, respectively.
These beamforming architectures are illustrated in Fig.~\ref{system model:sub2}.

\subsection{Channel Model and mmWave Downlink Communication}
Due to the limited diffraction ability of mmWave signals, we adopt a geometric mmWave channel model that consists of $C$ clusters, and each cluster contributes $L$ paths.
Let $\theta_{c,l}$, $\phi_{c,l}$ and $\alpha_{c,l}$ denote the angle of departure (AoD), angle of arrival (AoA), and the complex gain of the $l$-th path in the $c$-th cluster, respectively.
Then, the mmWave channel coefficients of the $u$-th UE, represented by $\mathbf{H}_u$, can be given by
\begin{equation}
\mathbf{H}_u=\sqrt{N_{\text{BS}}N_{\text{UE}}}\sum_{c=1}^{C}\sum_{l=1}^{L}\alpha_{c,l}\mathbf{a}_{\text{UE}}(\phi_{c,l})\mathbf{a}_{\text{BS}}^{H}(\theta_{c,l}),
\end{equation}
where 
\begin{align}
&\mathbf{a}_{\text{BS}}(\theta_{c,l})=\frac{1}{\sqrt{N_{BS}}}[1,e^{j\pi\sin(\theta_{c,l})},\dots,e^{j\pi(N_{BS}-1)\sin(\theta_{c,l})}]^T,\notag\\
&\mathbf{a}_{\text{UE}}(\phi_{c,l})=\frac{1}{\sqrt{N_{UE}}}[1,e^{j\pi\sin(\phi_{c,l})},\dots,e^{j\pi(N_{UE}-1)\sin(\phi_{c,l})}]^T,\notag
\end{align}
are the steering vectors corresponding to the AoD $\theta_{c,l}$ and AoA $\phi_{c,l}$, respectively.

\par
\begin{figure}[t]
\centering
\includegraphics[width=0.45\textwidth]{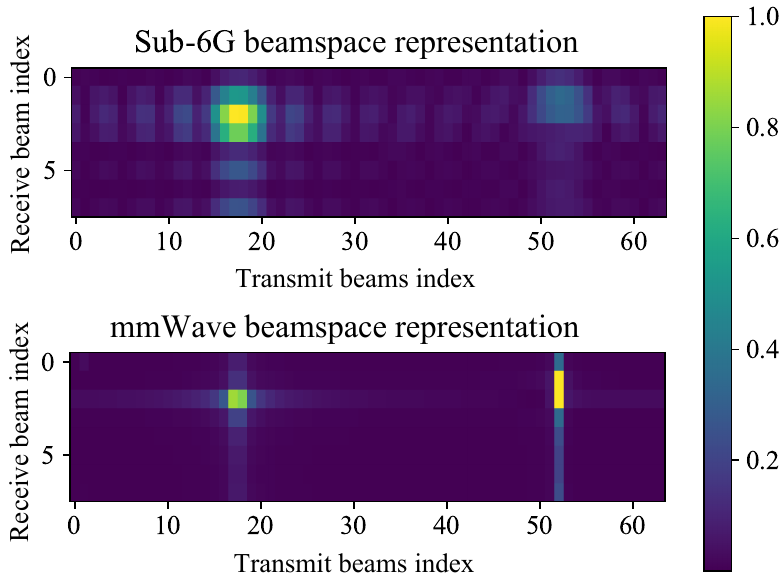}
\caption{Normalized beamspace representation of sub-6G channel and mmWave channel.}
\label{codebook_gain_matrix}
\end{figure}

 As for the sub-6G channel $\underline{{\mathbf H}_u}$, a similar geometric channel model with limited clusters is adopted.
As depicted in Fig.~\ref{codebook_gain_matrix}, the dominant paths of the sub-6G and mmWave channels partially overlap in the angular domain, however, their total number of clusters/paths, the channel gain of each path and other parameters are generally different \cite{ali2017millimeter}.
In the numerical simulations later, we adopt the DeepMIMO dataset \cite{Alkhateeb2019,Remcom} for network training, which is generated by running ray-tracing in sub-6G and mmWave frequencies.

\par For mmWave downlink communication, denote the transmit signal vector of at the BS by $\mathbf{s}\triangleq [s_1,\dots,s_K]^T$, where $s_{u}$ is the transmit signal to the $u$-th UE.
After transmit precoding, channel propagation and receive combining, the received signal at the $u$-th UE, represented by $y_u$, is given by
\begin{equation}
y_{u}=\mathbf{v}_{u}^{H}\mathbf{H}_{u}\mathbf{F}_{\text{RF}}\sum_{i = 1}^{K}{\mathbf{F}_{\text{BB}}}{[:,i]}s_i+\mathbf{v}_u^H \mathbf{n}_u, \label{receive signal}
\end{equation}
where $\mathbb{E}[\mathbf{s}\mathbf{s}^H]=\frac{P_{\text{t}}}{K}\mathbf{I}_K$ and $P_{\text{t}}$ denotes the total transmit power of the BS, $\mathbf{n}_u\in \mathcal{C}\mathcal{N}(0,\sigma_{n}^2\mathbf{I}_{N_{\text{UE}}})$ is the noise vector of the $u$-th UE with $\sigma_{n}^2$ being the noise power, $\mathbf{F}_{\text{RF}}\in\mathbb{C}^{N_{\text{BS}}\times N_{\text{RF}}}$ and $\mathbf{F}_{\text{BB}}\in \mathbb{C}^{N_{\text{RF}}\times N_{\text{RF}}}$ denote the analog precoder and digital precoder at the BS, respectively, and satisfy the power constraint $\left\lVert \mathbf{F}_{\text{RF}}\mathbf{F}_{\text{BB}}\right\rVert_{\ell_2}^2=K$, ${\mathbf v}_{u}\in {\mathbb C}^{N_{\text{UE}}\times 1}$ denotes the receive combiner of the $u$-th UE.
Specially, each ${\mathbf v}_u$ and each column of the analog precoder ${\mathbf F}_{\text{RF}}$ are selected from predefined DFT codebooks ${\mathbf Z}\in \mathbb{C}^{N_{\text{UE}}\times N_{\mathbf{Z}}}\triangleq [{\mathbf z}_1,\dots,{\mathbf z}_{N_{\text{Z}}}]$ and ${\mathbf W}\in \mathbb{C}^{N_{\text{BS}}\times N_{\mathbf{W}}}\triangleq [{\mathbf w}_1,\dots,{\mathbf w}_{N_{\text{W}}}]$, respectively, where $N_{\mathbf{Z}}$ and $N_{\mathbf{W}}$ denote the corresponding numbers of codewords.
Without loss of generality, we assume $N_{\mathbf{Z}}=N_{\text{UE}}$, $N_{\mathbf{W}}=N_{\text{BS}}$, and $N_{\text{RF}}=K$.
Furthermore, we define ${\mathbf h}_{u,\text{eff}}^H=\mathbf{v}_{u}^{H}\mathbf{H}_{u}\mathbf{F}_{\text{RF}}\in\mathbb{C}^{1\times N_{\text{RF}}}$ as the effective channel between the $u$-th UE and BS.
Accordingly, the received signal model in \eqref{receive signal} can be rewritten as
\begin{equation}
y_{u}={\mathbf h}_{u,\text{eff}}^H\sum_{i = 1}^{K}{\mathbf{F}_{\text{BB}}}{[:,i]}s_i+\mathbf{v}_u^H \mathbf{n}_u. \label{receive signal rewritten}
\end{equation}
As a result, the signal-to-interference-plus-noise ratio (SINR) of $y_{u}$, represented by $\gamma_{u}$, can be written as follows:
\begin{equation}
\gamma_u=\frac{\frac{P_{\text{t}}}{K}\left\lvert \mathbf{h}_{u,\text{eff}}^H{\mathbf{F}_{\text{BB}}}{[:,u]}\right\rvert^2}{\sum_{i\neq u}\frac{P_{\text{t}}}{K}\left\lvert \mathbf{h}_{u,\text{eff}}^H{\mathbf{F}_{\text{BB}}}{[:,i]}\right\rvert^2+\sigma_n^2\left\lVert \mathbf{v}_u\right\rVert^2_{\ell_2} }.
\end{equation}
Given $\gamma_u$, the spectrum efficiency of the $u$-th UE, denoted by $R_u$, is given by
\begin{equation}
R_u= \log_{2}(1+\gamma_u).\label{user spectrum efficiency}
\end{equation}
Then, the overall spectrum efficiency of the considered communication system, represented by $R_{\rm{sum}}$, is 
\begin{equation}
R_{\rm{sum}}=\sum_{u=1}^{K}R_u.\label{sum spectrum efficiency}
\end{equation}

\subsection{Problem formulation}
Our goal is to optimize the analog precoder and digital precoder at the BS $\{\mathbf{F}_{\text{RF}},\mathbf{F}_{\text{BB}}\}$ as well as the analog combiners at all UEs $\{{\mathbf v}_1,\dots,{\mathbf v}_K\}$ to maximize the sum spectrum efficiency $R_{\text{sum}}$ for the system considered. Mathematically, the following problem is formulated:
\begin{align}
\max_{\mathbf{F}_{\text{BB}},\mathbf{F}_{\text{RF}},\{{\mathbf v}_1,\dots,{\mathbf v}_K\}}&~R_{\rm{sum}} \label{max rate problem} \\
\text{s.t}\quad&~\mathbf{v}_i\in {\mathcal Z},\forall i\in\{1,\dots,K\}, \tag{\ref{max rate problem}{a}}\label{max rate problem:a}\\
&~{\mathbf{F}_{\text{RF}}}[:,j]\in {\mathcal W},\forall j\in\{1,\dots,N_{\text{RF}}\}, \tag{\ref{max rate problem}{b}}\label{max rate problem:b}\\
&\left\lVert \mathbf{F}_{\text{RF}}\mathbf{F}_{\text{BB}}\right\rVert_{\ell_2}^2=K,\tag{\ref{max rate problem}{c}}\label{max rate problem:c}&
\end{align}
where ${\mathcal Z}\triangleq \{{\mathbf z}_1,\dots,{\mathbf z}_{N_{\text{Z}}}\}$ and ${\mathcal W}\triangleq \{{\mathbf w}_1,\dots,{\mathbf w}_{N_{\text{W}}}\}$.

\par Traditionally, based on perfect mmWave CSI, a variety of algorithms have been proposed to solve problem \eqref{max rate problem}, employing optimization \cite{10187715} or deep learning techniques \cite{elbir2019hybrid,jinModelDrivenDeepLearning2023}.
However, the acquisition of accurate mmWave CSI requires a substantial number of pilots owing to the large number of antennas, even when using compressed sensing based algorithms \cite{Alkhateeb2014,alkhateeb2015compressed}.
Furthermore, some decoupled hybrid beamforming methods have been developed for mmWave systems with limited feedback \cite{alkhateeb2015limited,alkhateeb2016frequency}, but these methods require extensive beam training, particularly when the number of UEs is large.

\par Fortunately, in the dual-band system considered, the correlation between sub-6G and mmWave frequencies have been widely investigated and demonstrated in numerous works \cite{peter2016measurement,samimi20163}.
This correlation arises due to the presence of shared scatterers in the propagation environment, and thus the sub-6G channel and mmWave channel exhibit significant similarity in both time and spatial domains.
Capitalizing on the spatial congruence between sub-6G and mmWave frequencies presents a promising avenue to reduce the pilot signal overhead.
Motivated by these insights, our objective is to incorporate sub-6G channel information to support mmWave hybrid beamforming.
Additionally, we take a pragmatic approach by considering the imperfect sub-6G channel estimate in the design.

\par In particular, to estimate the sub-6G channels, each UE transmits uplink training pilot sequences to the BS.
Let $\underline{L_{\text{p}}}$ denote the training sequence length, and $\underline{{\mathbf S}_{\text{u,p}}} \in {\mathbb C}^{\underline{N_{\text{UE}}} \times \underline{L_{\text{p}}}} $ denote a pilot matrix with each row corresponding to a pilot sequence sent from a particular antenna of the $u$-th UE.  
Then, the received training signals at the BS are given by
\begin{equation}
\underline{{\mathbf Y}_{\text{p}}}=\sum\limits_{u=1}^{K}\underline{{\mathbf H}_u}^T\underline{{\mathbf S}_{\text{u,p}}}+\underline{{\mathbf N}_{\text{p}}},\label{sub_6G received signals}
\end{equation}
where $\underline{{\mathbf N}_{\text{p}}}\sim {\mathcal{CN}}(0,\underline{\sigma_n^2})$ is the noise matrix with independent and identically distributed (i.i.d.) noise elements, and $\underline{\sigma_n^2}$ denotes the noise power during pilot transmission.
In general, we should have $\underline{L_{\text{p}}} \geqslant K\underline{N_\text{UE}}$, to ensure the orthogonality between any two pilot sequences sent from different antennas of each user and from different users.
Here, we simply assume that $\underline{L_{\text{p}}}=K\underline{N_\text{UE}}$. 
With this assumption and defining $\underline{{\mathbf S}_{\text{p}}}=[\underline{{\mathbf S}_{\text{1,p}}}^T,\dots,\underline{{\mathbf S}_{\text{K,p}}}^T]^T\in {\mathbb C}^{{K\underline{N_{\text{UE}}}}\times \underline{L_{\text{p}}}}$, we can rewrite \eqref{sub_6G received signals} as 
\begin{equation}
\underline{{\mathbf Y}_{\text{p}}}=[\underline{{\mathbf H}_1}^T,\dots,\underline{{\mathbf H}_K}^T]\underline{{\mathbf S}_{\text{p}}}+\underline{{\mathbf N}_{\text{p}}},\label{sub_6G received signals rewritten}
\end{equation}
where $\underline{{\mathbf S}_{\text{p}}}$ satisfies $\underline{{\mathbf S}_{\text{p}}}^H\underline{{\mathbf S}_{\text{p}}}=\underline{P_{\text{s}}}{\mathbf I}_{\underline{L_{\text{p}}}}$ with $\underline{P_{\text{s}}}$ denoting the sub-6G pilot signal power.
Subsequently, the sub-6G channels can be obtained by resorting to the low-complexity least square estimation as 
\begin{equation}
[\underline{\widehat{\mathbf H}_1}^T,\dots,\underline{\widehat{\mathbf H}_K}^T]=\underline{{\mathbf Y}_{\text{p}}}~\underline{{\mathbf S}_{\text{p}}}^{-1}.\label{sub-6G channel estimation}
\end{equation}

\par In the subsequent sections, we shall focus on developing a framework to leverage such sub-6G channel estimate to facilitate mmWave hybrid beamforming.

\section{Sub-6G Information Aided Multi-User Hybrid Beamforming: Overall Design}
\label{Framework_of_Sub-6G_Information-Aided_Multi-User_Hybrid_Beamforming}
In this section, we first discuss the key issues of utilizing sub-6G information for mmWave hybrid beamforming and then present the overall design of the proposed SA-MUHBF framework.

\par To effectively integrate the sub-6G channel estimates into mmWave hybrid beamforming, two main issues arise: (1) identifying the relevant sub-6G channel information for mmWave beamforming, and (2) designing an efficient hybrid beamforming scheme based on the sub-6G channel information.

\par To address the first issue, we need to analyze the spatial congruence between sub-6G and mmWave channels. 
As illustrated in Fig.~\ref{codebook_gain_matrix}, although there is a discrepancy in the number of paths between sub-6G and mmWave channels, their dominant path angles exhibit a strong correlation, and the normalized gains of these paths are similar. 
This indicates that while directly reconstructing the full mmWave channel from sub-6G channel information is difficult, it is still feasible to predict its beamspace representation by capturing key features such as dominant angles and path gains.

\par To address the second issue, decoupling the optimization of analog and digital beamforming might offer an effective solution. 
One the one hand, the extracted beamspace information is particularly useful for analog beam selection at each user, enabling user separation in the spatial domain.
On the other hand, digital beamforming can be more reliably designed if additional limited in-band measurements are available. 

\par In light of the above discussions, we propose to construct the SA-MUHBF framework via three stages (illustrated in Fig.~\ref{diagram}), as follows:
\begin{figure*}[t]
\centering
\includegraphics[width=0.98\textwidth]{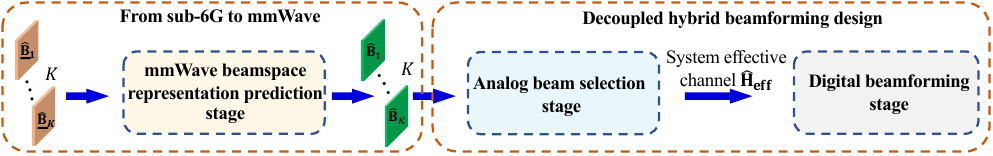}
\caption{The overall design of SA-MUHBF framework.}
\label{diagram}
\end{figure*}
\begin{itemize}
    \item mmWave beamspace representation prediction stage: Extract the spatial information from sub-6G channel estimates and then predict mmWave beamspace representations;
    \item Analog beam selection stage: Select the proper analog beams based on the predicted mmWave beamspace representations while accounting for inter-user interference coordination;
    \item Digital beamforming stage: Obtain an estimate of the effective multi-user channel $\widehat{\mathbf H}_{\text{eff}}$ in the digital domain after applying the analog beams selected from the previous stage, and then design an optimized digital beamforming based on $\widehat{\mathbf H}_{\text{eff}}$.
\end{itemize}

\section{Sub-6G Information Aided Multi-User Hybrid Beamforming: Deep-Learning Assisted Implementation}
\label{Architecture_and_Implementation_of_Sub-6G_Information-Aided_Multi-User_Hybrid_Beamforming}
We now proceed to present a deep-learning assisted implementation of the SA-MUHBF framework.
We first provide the detailed description of each stage, and then present the training procedure for the overall design.

\subsection{mmWave Beamspace Representation Prediction Stage}
\begin{figure}[t]
    \centering
    \includegraphics[width=0.45\textwidth]{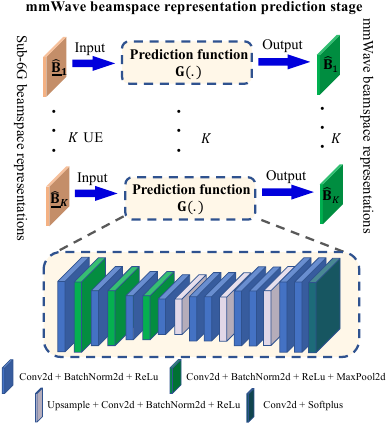}
    \caption{An illustration of the mmWave beamspace representation prediction stage.}
    \label{mmWave_information_prediction_stage}
    \end{figure}
This stage is aimed at extracting the mmWave spatial information from the sub-6G channel and then predicting the mmWave beamspace representation.
Taking the $u$-th UE as an example, to facilitate this extraction, we first convert the sub-6G channel estimate $\underline{\widehat{\mathbf H}_u}$ into a sub-6G beamspace representation $\underline{\widehat{\mathbf B}_u}$. 
This conversion is achieved by utilizing two oversampled DFT codebooks, denoted by $\underline{\mathbf Z}\in {\mathbb C}^{\underline{N_{\text{UE}}}\times N_{\text{Z}}}$ and $\underline{\mathbf W}\in {\mathbb C}^{\underline{N_{\text{BS}}}\times N_{\text{W}}}$, and the resultant $\underline{\widehat{\mathbf B}_u}$ is given by 
\begin{equation}
\underline{\widehat{\mathbf B}_u}={|\underline{\mathbf Z}^H\underline{\widehat{\mathbf H}_u}~\underline{\mathbf W}|}.\label{sub-6G channel2beamspace}
\end{equation}
Note that each element of $\underline{\widehat{\mathbf B}_u}$ indicates the likelihood of a strong path's presence in the corresponding beamspace bin.

\par Subsequently, a prediction function, represented by ${\mathbf G}(\cdot)$, is designed to map $\underline{\widehat{\mathbf B}_u}$ into a beamspace representation of the mmWave channel, i.e.,
\begin{equation}
\widehat{\mathbf B}_u={\mathbf G}(\underline{\widehat{\mathbf B}_u}).\label{mmWave beamspace representation prediction}
\end{equation}
Note that the predicted $\widehat{\mathbf B}_u \in {\mathbb R}_{+}^{N_{\text{Z}}\times N_{\text{W}}}$ is expected to match well with the true mmWave beamspace representation ${\mathbf B}_u$, which is given by ${\mathbf B}_u={|{\mathbf Z}^H{\mathbf H}_u{\mathbf W}|}$.

\par Towards this end and motivated by the recent success of CNNs in featrue learning, generative and reconstruction tasks \cite{6472238}, we propose to construct ${\mathbf G}(\cdot)$ based on 2D CNNs.
As illustrated in Fig.~\ref{mmWave_information_prediction_stage}, we adopt a stack structure of 2D CNN blocks, each of which consists of Conv2d, BatchNorm2d, ReLu, MaxPool2d and Upsample layers.
Specifically, the input $\widehat{\mathbf B}_u$ is first compressed via a group of 2D CNN blocks to generate essential features, which are then used to predict the mmWave beamspace representation through a second group of 2D CNN blocks.  
Moreover, to ensure that each element in the output $\widehat{\mathbf B}_u$ is non-negative, a Softplus activation function, i.e., $\text{Softplus}(x)=\log(1+e^x)$, is adopted in the end.

\subsection{Analog Beam Selection Stage}
\begin{figure}[t]
\centering
\includegraphics[width=0.48\textwidth]{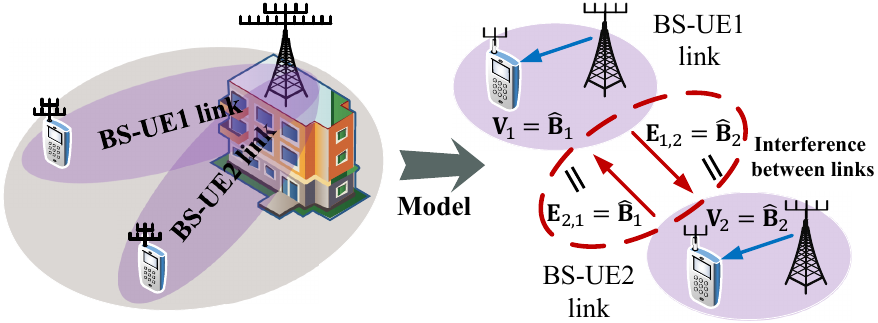}
\caption{Illustration of the interference graph.}
\label{simplified_interference_graph}
\end{figure}
Given the predicted mmWave beamspace representations $\{\widehat{\mathbf B}_1,\dots,\widehat{\mathbf B}_K\}$, this stage aims to perform analog beam selection for the BS and all UEs by assuming that ${\mathbf F}_{\text{BB}}={\mathbf I}_{N_{\text{RF}}}$.
Mathematically, we introduce a $N_{\text{Z}}\times N_{\text{W}}$ binary matrix $\tilde{\mathbf L}_u$ to represent the beam selection strategy for the $u$-th BS-UE link, where $\| \tilde{\mathbf L}_u\|_{\ell_0}=1$, and $\tilde{\mathbf L}_u[i,j]=1$ indicates that the $u$-th BS-UE link adopts ${\mathbf z}_i$ and ${\mathbf w}_j$ as the combiner and analog precoder, respectively.
Then, we shall design an efficient mapping function from $\{\widehat{\mathbf B}_1,\dots,\widehat{\mathbf B}_K\}$ to $\{\tilde{\mathbf L}_1,\dots,\tilde{\mathbf L}_K\}$. 

\par To cope with the multi-user interference, we propose a novel GNN-aided analog beam selection method.
In particular, we first construct a low-complexity interference graph based on $\{\widehat{\mathbf B}_1,\dots,\widehat{\mathbf B}_K\}$.
Then, we design a multi-layer GNN based beam selection model where a novel graph convolution layer is proposed to update the beam selection strategies by taking interference coordination into consideration. 

\subsubsection{Graph Modeling of the Multi-User mmWave System}
\label{graph_modeling_of_the_multi-user_mmWave_system}
Denote the interference graph by ${\mathcal G}({\mathcal V},{\mathcal E})$, where ${\mathcal V}$ and ${\mathcal E}$ represent the vertex set and edge set, respectively.
In ${\mathcal G}({\mathcal V},{\mathcal E})$, the $u$-th vertex is used to model the $u$-th BS-UE link, and the attribute of the $u$-th vertex, represented by ${\mathbf V}_u$, is defined as the effective channel strength of the $u$-th BS-UE link.
Additionally, the directed edge $(u,v)$ denotes the interference from the $u$-th link to the $v$-th link, and the attribute of the edge $(u,v)$, represented by ${\mathbf E}_{u,v}$, is defined as the effective interference channel strength from the $u$-th link to the $v$-th link.
In the considered multi-user system, due to the absence of perfect mmWave channel knowledge, we propose to construct ${\mathcal G}$ based on the obtained mmWave beamspace representation estimates $\{\widehat{\mathbf B}_1,\dots,\widehat{\mathbf B}_K\}$.
Taking the $u$-th vertex as an example, we set ${\mathbf V}_u=\widehat{\mathbf B}_u$ and ${\mathbf E}_{u,v}=\widehat{\mathbf B}_v$.
The resultant interference graph is shown on the right of Fig.~\ref{simplified_interference_graph}.
Since the attribute of edge $(u,v)$ coincides with the attribute of the $v$-th vertex, i.e., ${\mathbf E}_{u,v}={\mathbf V}_v=\widehat{\mathbf B}_v$, we can only store the vertex attribute when representing the graph in the processing.

\subsubsection{Multi-layer GNN Based Beam Selection Model}
With the constructed interference graph ${\mathcal G}$, we now propose a multi-layer GNN based beam selection model, represented by ${\mathbf S}(\cdot)$, to produce the analog beam selection strategies of all the BS-UE links.
Specifically, ${\mathbf S}(\cdot)$ is constructed by cascading $T$ graph convolution layers, as illustrated in Fig.~\ref{GNN_model}.
The inputs of each layer include the interference graph ${\mathcal G}$ and a set of beam selection probability matrices generated from the previous layer. 
Let ${\mathbf s}_t(\cdot)$ denote the $t$-th layer, and let ${\mathbf L}_u^t$ represent the beam selection probability matrix for the $u$-th BS-UE link that satisfies $\|{\mathbf L}^t_u\|_{\ell_1} = 1$.
Then, we have
\begin{equation}
\{{\mathbf L}_1^t,\dots,{\mathbf L}_K^t\}={\mathbf s}_t(\{{\mathbf L}_1^{t-1},\dots,{\mathbf L}_K^{t-1}\},{\mathcal G}).
\end{equation}
Note that when $t=1$, ${\mathbf L}_u^0$ is initialized by a normalized version of $\widehat{\mathbf B}_u$ from the beamspace prediction stage, i.e., 
\begin{equation}
{\mathbf L}_u^{0}=\frac{\widehat{\mathbf B}_u}{\|\widehat{\mathbf B}_u\|_{\ell_1}},~\forall~u\in \{1,\dots,K\}.\label{anglog beam selection initialization}
\end{equation}

\par After the processing of $T$ graph convolution layers (or equivalently $T$ iterations), the produced beam selection probability matrix ${\mathbf L}_u^T$ is further mapped into a binary matrix $\tilde{{\mathbf L}}_u$ according to the following rule:
\begin{equation}
\tilde{\mathbf L}_u[i,j]=
\begin{cases}
1,~i=i_u^{\star},~j=j_u^{\star},\\
0,~\text{otherwise}
\end{cases},~\forall u\in \{1,\dots,K\},
\end{equation}
where $(i_u^{\star},j_u^{\star})=\arg \max_{i,j} {\mathbf L}_u^T[i,j]$ contains the row and column indices corresponding to the largest element in ${\mathbf L}_u^T$. 
Then, the analog precoder and combiner for the $u$-th BS-UE link are set as
\begin{gather}
{\mathbf F}_{\text{RF}}[:,u]={\mathbf w}_{j_u^{\star}},~{\mathbf v}_u={\mathbf z}_{i_u^{\star}},~\forall~u \in \{1,\dots,K\}.\label{analog beamforming vectors}
\end{gather}

\par Next, we present the architecture of the graph convolution layer in detail. 

\subsubsection{Architecture of the Graph Convolution Layer}
\label{Architecture of Graph Convolution Layer}
\begin{figure*}[t]
\centering
\includegraphics[width=0.96\textwidth]{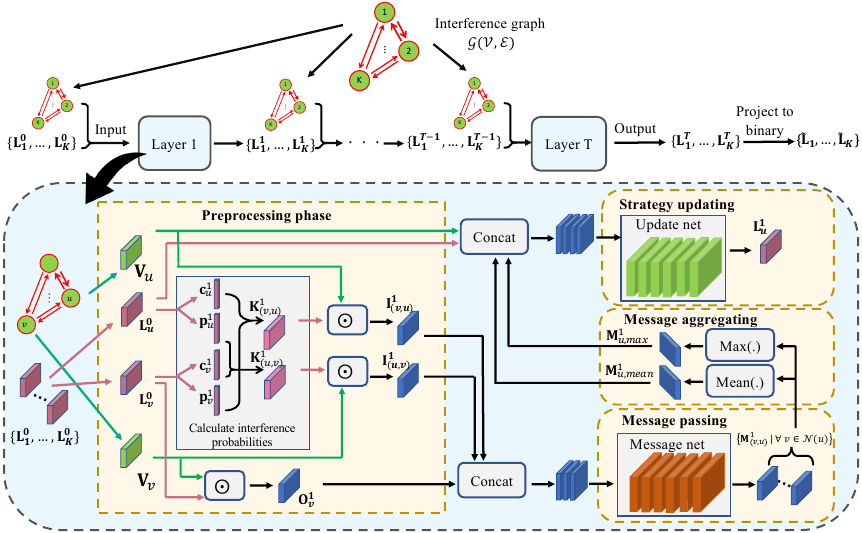}
\caption{Illustration of our proposed multi-layer GNN model.}
\label{GNN_model}
\end{figure*}
Aiming to improve the analog beam selection quality, we propose a novel graph convolution layer, which incorporates a preprocessing phase for the inputs and a 2D CNN based graph convolution for the preprocessed features, as illustrated in Fig.~\ref{GNN_model}.
In the preprocessing phase, both the effective signal strength and interference strength among different BS-UE links are explicitly modeled.
Specifically, taking the $t$-th layer as an example, we first calculate the probabilities of different analog combiners at the UEs and different analog precoders at the BS, represented by $\{{\mathbf c}_1^t,\dots,{\mathbf c}_K^t\}$ and $\{{\mathbf p}_1^t,\dots,{\mathbf p}_K^t\}$, respectively, by taking $\{{\mathbf L}^{t-1}_1,\dots,{\mathbf L}^{t-1}_K\}$ and ${\mathcal G}$ as inputs, i.e.,
\begin{align}
{\mathbf c}_u^t[i]&= \sum_{j=1}^{N_{\mathbf Z}}{\mathbf L}_u^t[i,j],~\forall~i \in \{1,\dots,N_{\mathbf Z}\},~\forall~u \in \{1,\dots,K\},\notag\\
{\mathbf p}_u^t[j]&= \sum_{i=1}^{N_{\mathbf W}}{\mathbf L}_u^t[i,j],~\forall~j \in \{1,\dots,N_{\mathbf W}\},~\forall~u \in \{1,\dots,K\},\notag
\end{align}
where ${\mathbf c}_u^t[i]$ and ${\mathbf p}_u^t[j]$ represent the probabilities of the $u$-th BS-UE link selecting ${\mathbf z}_i$ and ${\mathbf w}_j$ as the combiner and analog precoder, respectively.
Subsequently, with $\{{\mathbf c}_1^t,\dots,{\mathbf c}_K^t\}$ and $\{{\mathbf p}_1^t,\dots,{\mathbf p}_K^t\}$, we obtain the interference probabilities among different links, represented by ${\mathbf K}_{(u,v)}^t$, as follows:
\begin{equation}
{\mathbf K}_{(u,v)}^t={\mathbf c}_v^t({{\mathbf p}_u^t})^T,~\forall~u,~v\in \{1,\dots,K\},~u\neq v,
\end{equation}
where ${\mathbf K}_{(u,v)}^t\in {\mathbb R}_{+}^{N_{\text{Z}}\times N_{\text{W}}}$ and $\| {\mathbf K}_{(u,v)}^t \|_{\ell_1}=1$.
Then, ${\mathbf K}_{(u,v)}^t$ is further converted to a matrix ${\mathbf I}^t(u,v) \in {\mathbb R}_{+}^{N_{\text{Z}}\times N_{\text{W}}}$ to capture the potential interference strength of the $u$-th BS-UE link to the $v$-th BS-UE link, via
\begin{equation}
{\mathbf I}_{(u,v)}^t={\mathbf K}_{(u,v)}^t\odot {\mathbf V}_v,~\forall~u,~v\in \{1,\dots,K\},~u\neq v.
\end{equation} 
Additionally, the potential effective signal strength of the $u$-th BS-UE link, denoted by ${\mathbf O}_u^t \in {\mathbb R}_{+}^{N_{\text{Z}}\times N_{\text{W}}}$ can be obtained as 
\begin{equation}
{\mathbf O}_u^t={\mathbf L}_u^{t-1}\odot {\mathbf V}_u,~\forall~u\in \{1,\dots,K\}.
\end{equation}
The above preprocessing effectively fuses the predicted beamspace representation and beam selection probabilities generated from the previous graph convolution layer.

\par Next, the obtained 2D matrices $\{{\mathbf O}_u^t|\forall u\in \{1,\dots,K\}\}$ and $\{{\mathbf I}_{(u,v)}^t|~\forall ~u,~v\in \{1,\dots,K\},~u\neq v\}$ are further processed by graph convolution, which involves three phases, i.e., message passing, message aggregation and strategy updating.

\paragraph*{Message Passing Phase}
Specifically, taking the $u$-th vertex and its neighboring vertex $v$ as an example, we employ a 2D CNN structure, denoted by $\text{CNN1}(\cdot)$, to extract features from the mutual potential interference strength $\{{\mathbf I}_{(u,v)}^t,{\mathbf I}_{(v,u)}^t\}$ and the effective signal strength of the $u$-th link ${\mathbf O}_u^t$ as the message.
In particular, the message from the $u$-th vertex to the $v$-th vertex is denoted by ${\mathbf M}_{(u,v)}^t\in {\mathbb R}_{+}^{N_{\text{Z}}\times N_{\text{W}}}$ and is calculated as 
\begin{equation}
{\mathbf M}_{(u,v)}^t=\text{CNN1}\left(\text{CONCAT}({\mathbf O}_u^t,{\mathbf I}_{(u,v)}^t,{\mathbf I}_{(v,u)}^t)\right), \label{message passing}
\end{equation}
where $\text{CONCAT}(\cdot)$ concatenates the input matrices in the $0$-th dimension.
Each vertex adopts the same operation as in \eqref{message passing} to produce its messages and subsequently sends them to its neighboring vertices along the directed edges.

\paragraph*{Message Aggregating Phase} In this phase, each vertex conducts message aggregation over its received messages.
For instance, denoting the received messages of the $u$-th vertex by $\{{\mathbf M}_{(u,v)}~|~\forall~v\in {\mathcal N}(u)\}$, then the message aggregation is implemented through
\begin{align}
{\mathbf M}_{u,\text{max}}^t&=\text{MAX}(\{{\mathbf M}_{(v,u)}^t~|~\forall~v\in {\mathcal N}(u)\}),\notag\\ 
{\mathbf M}_{u,\text{mean}}^t&=\text{MEAN}(\{{\mathbf M}_{(v,u)}^t~|~\forall~v\in {\mathcal N}(u)\}),
\end{align}
where $\text{MAX}(\cdot)$ selects the maximum value of the input matrices in an element-wise manner, i.e.,
\begin{gather}
{\mathbf M}_{u,\text{max}}^t[i,j]=\max\left(\{{\mathbf M}_{(v,u)}^t[i,j]~|~\forall~v\in {\mathcal N}(u)\}\right),\notag\\
 \forall~i \in \{1,\dots,N_{\mathbf Z}\},\forall~j \in \{1,\dots,N_{\mathbf W}\}.
\end{gather}
Similarly, the operation $\text{MEAN}(\cdot)$ returns the mean value of the input matrices in an element-wise manner, i.e.,
\begin{gather}
{\mathbf M}_{u,\text{mean}}^t[i,j]=\frac{1}{\text{card}({\mathcal N}(u))}\sum\limits_{v\in {\mathcal N}(u)}{\mathbf M}_{(v,u)}^t[i,j],\notag\\
\forall~i \in \{1,\dots,N_{\mathbf Z}\},\forall~j \in \{1,\dots,N_{\mathbf W}\}.
\end{gather}

\paragraph*{Strategy Updating Phase} With ${\mathbf M}_{u,\text{max}}^t$, ${\mathbf M}_{u,\text{mean}}^t$, ${\mathbf L}_u^{t-1}$, and ${\mathbf V}_u$, we employ a 2D CNN structure again to combine them and then produce the updated beam pair selection probability ${\mathbf L}_u^{t}$ as 
\begin{equation}
{\mathbf L}_u^t=\text{CNN2}\left(\text{CONCAT}({\mathbf M}_{u,\text{max}}^t,{\mathbf M}_{u,\text{mean}}^t, {\mathbf L}_u^{t-1},{\mathbf V}_u)\right). \notag
\end{equation}

\par Overall, it can be seen that the proposed analog beam selection network design follows the ``learning and optimization'' paradigm \cite{he2021overview}.
\textcolor{black}{In particular, the GNN proposed comprises multiple graph convolution layers, each utilizing the interference graph and a set of beam selection probability matrices generated by the previous
layer as input. These layers iteratively update the beam selection strategies, leading to
a learning process where the strategies are progressively refined. Additionally, each graph convolutional layer includes a customized 2D CNN-based preprocessing phase for the inputs.}
Such customized design is different from several state-of-the-art GNNs, such as the TransformerConv GNN \cite{shiMaskedLabelPrediction2021a} and the GNN in GBLinks~\cite{he2022gblinks}, which do not preprocess the inputs and utilize either 1D CNNs or MLPs for graph convolution. 
Therefore, ${\mathbf S}(\cdot)$ is envisioned to exhibit superior interpretability and more robust performance, as will be validated in Section~\ref{Numerical_Results}.

\subsection{Digital Beamforming Stage}
Upon obtaining the analog precoder at BS and combiners at UEs from the previous stage, the digital precoder at the BS are then optimized in this stage. 
To this end, we propose to first estimate the effective BS-UE channel ${\mathbf H}_{\text{eff}}$ in the digital domain with the given analog precoder and combiners and then optimize the digital precoder based on the estimated $\widehat{\mathbf H}_{\text{eff}}$.

Specifically, to obtain $\widehat{\mathbf H}_{\text{eff}}$, each UE transmits uplink training pilot sequences to the BS.
Let $L_{\text{p}}$ represent the training pilot sequence length, and let ${\mathbf s}_{\text{u,p}}^T\in {\mathbb C}^{1\times L_{\text{p}}}$ denote the pilot vector of the $u$-th UE.
Then, the received training signals at the BS is given by
\begin{align}
{\mathbf Y}_{\text{p}}&=\sum\limits_{u=1}^{K}{\mathbf F}_{\text{RF}}^T{\mathbf H}_{u}^T{\mathbf v}_{u}^*{\mathbf s}_{\text{u,p}}^T+{\mathbf F}_{\text{RF}}^T{\mathbf N}_{\text{p}},\notag\\
&=\sum\limits_{u=1}^{K}{\mathbf h}_{\text{u,eff}}{\mathbf s}_{\text{u,p}}^T+{\mathbf F}_{\text{RF}}^T{\mathbf N}_{\text{p}},\label{mmWave pilot recieved signal}
\end{align}
where ${\mathbf N}_{\text{p}}\sim \mathcal{CN}(0,\sigma_n^2{\mathbf I}_{N_{\text{BS}}})$ is the noise matrix with i.i.d. noise elements, and $\sigma_n^2$ denotes the channel noise power.
In general, we should have $L_{\text{p}}\geqslant K$ to ensure the orthogonality of any two pilot sequences of different UEs, and here we assume that $L_{\text{p}}= K$.
With this assumption and by defining ${\mathbf S}_{\text{p}}=[{\mathbf s}_{\text{u,p}}^T,\dots,{\mathbf s}_{\text{u,p}}^T]^T\in {\mathbb C}^{K\times L_{\text{p}}}$, we can rewrite \eqref{mmWave pilot recieved signal} as
\begin{align}
{\mathbf Y}_{\text{p}}&=[{\mathbf h}_{\text{1,eff}},\dots,{\mathbf h}_{\text{K,eff}}]{\mathbf S}_{\text{p}}+{\mathbf F}_{\text{RF}}^T{\mathbf N}_{\text{p}},\notag\\
&={\mathbf H}_{\text{eff}}^T{\mathbf S}_{\text{p}}+{\mathbf F}_{\text{RF}}^T{\mathbf N}_{\text{p}},
\end{align}
where ${\mathbf S}_{\text{p}}$ satisfies ${\mathbf S}_{\text{p}}^H{\mathbf S}_{\text{p}}=P_{\text{s}}{\mathbf I}_{L_{\text{p}}}$, and $P_{\text{s}}$ represents the mmWave training pilot signal power.
As a result, the effective channel ${\mathbf H}_{\text{eff}}$ can be estimated as 
\begin{equation}
\widehat{\mathbf H}_{\text{eff}}=({\mathbf Y}_{\text{p}}{\mathbf S}_{\text{p}}^{-1})^T.\label{mmWave effective channels estimation}
\end{equation}

\par Based on $\widehat{\mathbf H}_{\text{eff}}\triangleq [\hat{\mathbf h}_{\text{1,eff}}^*,\dots,\hat{\mathbf h}_{\text{K,eff}}^*]$, the LMMSE digital beamforming scheme proposed in \cite{nguyen2014mmse} is adopted for inter-user interference coordination in the digital domain. 
\textcolor{black}{Specifically, the LMMSE digital precoder ${\mathbf F}_{\text{BB}}$ is derived as
\begin{equation}
{\mathbf{F}}_{\text{BB}}=\frac{\sqrt{K}\tilde{\mathbf{F}}_{\text{BB}}}{\lVert\mathbf{F}_{\text{RF}}\tilde{\mathbf{F}}_{\text{BB}}\lVert_{\ell_2}},\label{mmWave digital beamforming scheme}
\end{equation}
where $\tilde{\mathbf{F}}_{\text{BB}}$ is defined as 
\begin{equation}
{\tilde{\mathbf{F}}_{\text{BB}}}=\widehat{\mathbf{H}}_{\text{eff}}^H\left( \widehat{\mathbf{H}}_{\text{eff}}\widehat{\mathbf{H}}_{\text{eff}}^H+ \frac{K\sigma_{n}^2}{P_t}\mathbf{I}_{\text{K}} \right)^{-1}\mathbf{G},
\end{equation}
with $\mathbf{G}=\text{diag}\left\{ \| \hat{\mathbf{h}}_{1,\text{eff}} \|_{\ell_2},\dots, \|\hat{\mathbf{h}}_{K,\text{eff}} \|_{\ell_2}\right\}$, and the normalized operation in \eqref{mmWave digital beamforming scheme} is to ensure the power constraint \eqref{max rate problem:c} is met.}
\textcolor{black}{The entire implementation of SA-MUHBF is summarized in Algorithm~\ref{SA-MUHBF algorithm}.}
\begin{algorithm}[!ht]
    \small
    \renewcommand{\algorithmicrequire}{\textbf{Input:}}
    \renewcommand{\algorithmicensure}{\textbf{Output:}}
    \caption{The SA-MUHBF Proposed}
    \label{SA-MUHBF algorithm}
    
    \begin{algorithmic}[1]
        \Require sub-6G channel estimates of $K$ UEs $\underline{\widehat{\mathbf H}_1},\dots,\underline{\widehat{\mathbf H}_K}$; % input
        \Ensure precoders at BS ${\mathbf F}_{\text{RF}},{\mathbf F}_{\text{BB}}$, and combiners at $K$ UEs ${\mathbf v}_1,\dots,{\mathbf v}_K$; % output
        
        \Statex /* {\bf mmWave beamspace representation prediction} */ 
        \State Convert sub-6G channel estimates to its beamspace representations $\underline{\widehat{\mathbf B}_1},\dots,\underline{\widehat{\mathbf B}_K}$ through \eqref{sub-6G channel2beamspace}.
        \State Predict mmWave beamspace representations $\widehat{\mathbf B}_1,\dots,\widehat{\mathbf B}_K$ using a CNN model ${\mathbf G}(\cdot)$: $\widehat{\mathbf B}_u={\mathbf G}(\underline{\widehat{\mathbf B}_u}), \forall u=1,\dots,K.$
        
        \Statex /* {\bf Analog beam selection} */ 
        \State Construct wireless interference graph ${\mathcal G}(\mathcal{V},\mathcal{E})$ based on $\widehat{\mathbf B}_1,\dots,\widehat{\mathbf B}_K$ according to Sec.~\ref{graph_modeling_of_the_multi-user_mmWave_system};
        \Statex Initialize the analog beam selection strategies ${\mathbf L}_1^{0},\dots,{\mathbf L}_K^{0}$ through \eqref{anglog beam selection initialization}.
        \State Iteratively update the analog beam selection strategies through GNN model ${\mathbf S}(\cdot)$, and output the final strategies $\tilde{\mathbf L}_1,\dots,\tilde{\mathbf L}_K$.
        \State Select ${\mathbf F}_{\text{RF}},{\mathbf v}_1,\dots,{\mathbf v}_K,$ from their codebooks ${\mathbf Z},{\mathbf W}$ via \eqref{analog beamforming vectors}.

        \Statex /* {\bf Digital beamforming} */ 
        \State Estimate the effective channels $\widehat{\mathbf H}_{\text{eff}}$ via \eqref{mmWave pilot recieved signal}-\eqref{mmWave effective channels estimation}.
        \State Calculate the LMMSE digital precoder $\mathbf{F}_{\text{BB}}$ through \eqref{mmWave digital beamforming scheme}.
        
        \State \Return ${\mathbf F}_{\text{RF}},{\mathbf F}_{\text{BB}},{\mathbf v}_1,\dots,{\mathbf v}_K$.
 \end{algorithmic}
\end{algorithm}

\subsection{Training of SA-MUHBF}
To train the proposed SA-MUHBF, a two-step approach is employed, where we first train the prediction network ${\mathbf G}(\cdot)$, and then train the GNN ${\mathbf S}(\cdot)$ by freezing the parameters of ${\mathbf G}(\cdot)$.
\subsubsection{Training of the Prediction Network ${\mathbf G}(\cdot)$}
The training process of ${\mathbf G}(\cdot)$ is conducted in a supervised manner.
The loss function is defined as the normalized mean square error (NMSE) between the predicted mmWave beamspace representation $\widehat{\mathbf B}$ and the true mmWave beamspace representation ${\mathbf B}$, i.e., 
\begin{equation}
l_1=\frac{\| \widehat{\mathbf B}-{\mathbf B}\|_{\ell_2}}{\left\lVert{\mathbf B}\right\lVert_{\ell_2}}.\label{l1}
\end{equation}
Minimizing $l_1$ encourages the convergence of the predicted $\widehat{\mathbf B}$ towards the true ${\mathbf B}$.

\subsubsection{Training of the Multi-Layer GNN Model ${\mathbf S}(\cdot)$}
\par Once ${\mathbf G}(\cdot)$ is trained, we proceed to train the GNN model ${\mathbf S}(\cdot)$ in an unsupervised manner by fixing the parameters of ${\mathbf G}(\cdot)$.
The goal of training ${\mathbf S}(\cdot)$ is to generate beam pair selection strategies $\{{\mathbf L}_1,\dots,{\mathbf L}_K\}$ that maximize the sum spectrum efficiency defined in \eqref{sum spectrum efficiency}.
However, it is noted that the projection operation at the end of ${\mathbf S}(\cdot)$ is not differentiable, which prevents the back propagation in the training. 
To tackle this problem, we propose a novel beam selection probability based loss function.
To improve the convergence of the multi-layer GNN, instead of only using the beam selection probabilities generated by the last layer, we propose to aggregate the beam selection probabilities at all the layers, i.e.,
{
\begin{equation}
\bar{\mathbf L}_u=\alpha_1{\mathbf L}_u^1+\dots+\alpha_T{\mathbf L}_u^T,~\forall~u\in \{1,\dots,K\},
\end{equation}}
where $\alpha_1,\dots,\alpha_T$ denote the combining weights that satisfy $\sum\nolimits_{t=1}^{T}\alpha_t=1$ and $\alpha_1<\dots<\alpha_T$. Moreover, based on $\{\bar{\mathbf L}_1,\dots,\bar{\mathbf L}_K\}$, we can obtain the aggregated interference probability, represented by $\{\bar{\mathbf K}_{(u,v)}|~\forall ~u,~v\in \{1,\dots,K\},~u\neq v\}$, as follows:
{
\begin{align}
&\bar{\mathbf K}_{(u,v)}=\bar{\mathbf c}_v{\bar{\mathbf p}_u}^T,~\forall~u,~v\in \{1,\dots,K\},~u\neq v,\notag\\
&\bar{\mathbf c}_v[i]= \sum_{j=1}^{N_{\mathbf Z}}\bar{\mathbf L}_v[i,j],\forall i \in \{1,\dots,N_{\mathbf Z}\}, \forall v \in \{1,\dots,K\},\notag\\
&\bar{\mathbf p}_u[j]= \sum_{i=1}^{N_{\mathbf W}}\bar{\mathbf L}_u[i,j],\forall j \in \{1,\dots,N_{\mathbf W}\}, \forall u \in \{1,\dots,K\}.\notag
\end{align}}

\par In this way, we can construct a surrogate function of the original spectrum efficiency in \eqref{sum spectrum efficiency} as
\begin{gather}
\bar{R}_u=\log_2\left(1+\frac{\sum \limits_{i,j}\bar{\mathbf L}_u[i,j]\frac{P_{\text{t}}}{K}|{\mathbf z}_i^H{\mathbf H}_u{\mathbf w}_j|^2}{\sum \limits_{v\neq u,i,j}\bar{\mathbf K}_{(v,u)}[i,j]\frac{P_{\text{t}}}{K}\left\rvert {\mathbf z}_i^H{\mathbf H}_u{\mathbf w}_j\right\rvert^2+\sigma_n^2 }\right),\notag\\
\forall~u\in\{1,\dots,K\},
\end{gather}
and form the following loss function for training ${\mathbf S}(\cdot)$:
\begin{equation}
    l_2=-\sum_{u=1}^{K}\bar{R}_u.
\end{equation}

\section{Numerical Results}
\label{Numerical_Results}
In this section, we conduct extensive numerical simulations to evaluate the proposed SA-MUHBF based on the DeepMIMO dataset \cite{Alkhateeb2019,Remcom}.
We first introduce the simulation setup and training parameters.
Subsequently, we compare the performance of SA-MUHBF with various baselines, verifying its performance gain and generalization capability under various system configurations.

\subsection{Simulation Setup and Training Parameters}
\begin{figure}[t]
\centering
\includegraphics[width=0.45\textwidth]{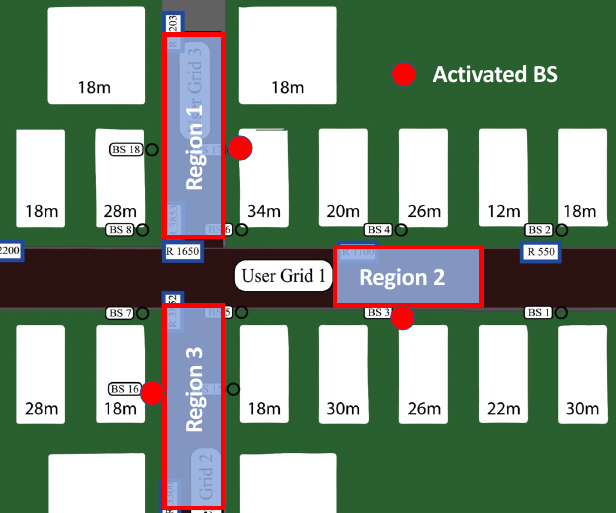}
\caption{Top view of the ``O1'' scenario and the regions used in our simulation.}
\label{O1 top}
\end{figure}
Consider the outdoor ``O1'' scenario of the DeepMIMO dataset, the top view of which is depicted in Fig.~\ref{O1 top}.
Three regions, i.e., BS 17 (R4000-R5200), BS 3 (R750-R1300), and BS 15 (R3300-R3800), as depicted in Fig.~\ref{O1 top}, are selected for the subsequent simulation.
The heights of the three activated BSs are 15m, while the UEs in these regions are at the height of 1.5m and are uniformly distributed.
The relevant parameters for the dual-band system considered are provided in Table~\ref{detailed parameters}. 
\begin{table}[t]
\renewcommand{\arraystretch}{2.0}
\caption{Simulation Parameters}
\label{detailed parameters}
\centering
\setlength\tabcolsep{0.5mm}{
\begin{tabular}{c|c|c}
\toprule
Notations                                   & Parameters                       & Values    \\ 
\midrule
$f_c$,~$\underline{f_c}$                    & \makecell{Operating frequency of \\ mmWave/sub-6G system  (GHz)}        & 28, 3.5 \\ \hline
$W$,~$\underline{W}$& \makecell{Bandwidth of \\mmWave/sub-6G system (GHz) }& 0.5, 0.02\\ \hline
$N_{\text{BS}}$,~$\underline{N_{\text{BS}}}$ & \makecell{Number of mmWave/sub-6G \\antennas at the BS} & 64, 16  \\ \hline
$N_{\text{UE}}$,~$\underline{N_{\text{UE}}}$ & \makecell{Number of mmWave/sub-6G \\antennas at each UE} &  8, 4    \\ \hline
$P_{\text{t}}$& \makecell{Total power of \\mmWave downlink (dBm)}& 40\\ \hline
$P_{\text{s}}$,~$\underline{P_{\text{s}}}$ & \makecell{Power of mmWave/sub-6G \\pilot signals (dBm) }& 30, 20\\ \hline
$\sigma_n^2$,~$\underline{\sigma_n^2}$ & \makecell{Noise power of \\mmWave/sub-6G system (dBm) }& \makecell{$-173.8 + 90 + 10\log_{10}(W)$\\$-173.8 + 90 + 10\log_{10}(\underline{W})$}\\
\bottomrule
\end{tabular}}
\end{table}

\par Besides, the network parameters of the beamspace representation prediction network ${\mathbf G}(\cdot)$ are provided in the left column of Table~\ref{network parameters}.
To train ${\mathbf G}(\cdot)$, we first collect samples from Region 1.
\textcolor{black}{Specifically, after running ray-tracing in the sub-6G and mmWave frequencies, we first obtain a total number of 200,000 samples, each of which is represented by $\{\underline{\mathbf H},{\mathbf H}\}$.
Considering the imperfect sub-6G channel estimation, given a sub-6G pilot power $\underline{P_{\text{s}}}$, we further acquire $\widehat{\underline{\mathbf H}}$ via \eqref{sub_6G received signals}-\eqref{sub-6G channel estimation}.
Following that, $\{\widehat{\underline{\mathbf H}},{\mathbf H}\}$ are converted into their beamspace representations $\{\widehat{\underline{\mathbf B}}, {\mathbf B}\}$, respectively.
Then, the converted samples are randomly divided into a training dataset, a validation dataset, and a testing dataset with a partition ratio of 0.4:0.1:0.5, and the batchsize $b_{\text{G}}$ is fixed at 128.}
The initial learning rate of ${\mathbf G}(\cdot)$, denoted by $\delta_{\text{G}}$, is set to $1\times 10^{-4}$, and is scheduled via the ``ReduceLROnPlateau'' learning rate scheduler \cite{noauthor_reducelronplateau_nodate}, which automatically reduces the learning rate when the model's performance on the validation set ceases to improve or demonstrates only marginal improvements over a specified number of training epochs.

\par As for the analog beam selection network ${\mathbf S}(\cdot)$, we consider a GNN model consisting of $T=3$ graph convolution layers, and the corresponding combining weights $\alpha_1$, $\alpha_2$, and $\alpha_3$ are set to 0.1, 0.2, and 0.7, respectively.
Each layer of ${\mathbf S}(\cdot)$ has the same network architecture, and the detailed parameters are provided in Table~\ref{network parameters}.
To train ${\mathbf S}(\cdot)$, we first collect a total number of 24,000 multi-user samples from Region 1, each of which includes the sub-6G channel estimates and mmWave channels of all the $K$ UEs, i.e., $\{\underline{\widehat{\mathbf H}_1},{\mathbf H}_1,\dots,\underline{\widehat{\mathbf H}_K},{\mathbf H}_K\}$.
These samples are fed into the trained ${\mathbf G}(\cdot)$ to generate the predicted mmWave beamspace representations as new data samples for ${\mathbf S}(\cdot)$, which are represented by $\{\underline{\widehat{\mathbf B}_1},{\mathbf H}_1,\dots,\underline{\widehat{\mathbf B}_K},{\mathbf H}_K\}$.
The obtained data samples are further divided into a training dataset, a validation dataset, and a testing dataset with a partition ratio of 0.4:0.1:0.5, and the batchsize $b_{\text{S}}$ is also fixed at 128.
The initial learning rate of ${\mathbf S}(\cdot)$, denoted by $\delta_{\text{S}}$, is also set to $1\times 10^{-4}$, which will be automatically adjusted through the ``ReduceLROnPlateau'' learning rate scheduler, similar to the training of ${\mathbf G}(\cdot)$. 

\begin{table}[t]
\caption{Input/Output Channels of Each Block in \\${\mathbf G}(\cdot)$, $\text{CNN1}(\cdot)$ and $\text{CNN2}(\cdot)$}
\renewcommand{\arraystretch}{1.2}
\label{network parameters}
\centering
\setlength\tabcolsep{2mm}{
\begin{tabular}{c|c|c}
\toprule
${\mathbf G}(\cdot)$ &$\text{CNN1}(\cdot)$ & $\text{CNN2}(\cdot)$\\
\midrule
(1,~64)&(3,~12)&(4,~12)\\
(64,~64)&(12,~24)&(12,~24)\\
(64,~128)&(24,~12)&(24,~12)\\
(128,~128)&(12,~6)&(12,~6)\\
(128,~256)&(6,~1)&(6,~1)\\
(256,~256)$\times 3$ &&\\
(256,~128)& &\\
(128,~128)$\times 2$&&\\
(128,~64) & &\\
(64,~64)$\times 4$&&\\
(64,~1)& &\\
\bottomrule
\end{tabular}}
\end{table}

\subsection{Baselines}
For performance comparison, we first introduce two baselines from \cite{maschiettiCoordinatedBeamSelection2019} as follows:
\begin{itemize}
    \item {\bf Uncoordinated method}: Define $b_{\text{r}}=\frac{N_{\text{UE}}}{\underline{N_{\text{UE}}}}$ and $b_{\text{c}}=\frac{N_{\text{BS}}}{\underline{N_{\text{BS}}}}$ to account for the angular resolution difference between sub-6G and mmWave channels.
    For the $u$-th ($1\leqslant  u\leqslant K$) BS-UE link, given the beamspace representation of sub-6G channel estimate $\underline{\widehat{\mathbf B}_u}$, the method first search a subset of candidate mmWave beams ${\mathcal C}_u = \{(\mathbf{z}_{i}, \mathbf{w}_{j}) | r_u^\star \leqslant i \leqslant r_u^\star + b_{\text{r}} - 1, c_u^\star \leqslant j \leqslant c_u^\star + b_{\text{c}} - 1\}$, with $\text{card}({\mathcal C}_u) = b_r b_c$, where $r_u^\star$ and $c_u^\star$ are index parameters determined by
    \begin{equation}
    (r_u^\star,c_u^\star)=\arg\max_{r,c}\frac{1}{b_{\text{r}}b_{\text{c}}}\sum\limits_{i=r}^{r+b_{\text{r}}-1}\sum\limits_{j=c}^{c+b_{\text{c}}-1}|\underline{\widehat{\mathbf B}_u}[i,j]|. \notag
    \end{equation}
    Next, we identify the analog beam pair for the $u$-th BS-UE link $(i_u^{\star},j_u^{\star})$ by searching the beam pair from $\mathcal C_u$ that leads to the highest measured output energy through mmWave beam training.
    As for digital beamforming, LMMSE precoder is formed based on estimate of effective channel in digital domain after applying the analog beams selected for each BS-UE link, as that in SA-MUHBF.   
    \item {\bf Coordinated method}: \textcolor{black}{This method is a sequential mmWave analog beam selection given an arbitrary order of BS-UE links.
    For the first BS-UE link, we first search for a subset of candidate mmWave beams ${\mathcal C}_1 = \{(\mathbf{z}_{i}, \mathbf{w}_{j}) | r_1^\star \leqslant i \leqslant r_1^\star + b_{\text{r}} - 1, c_1^\star \leqslant j \leqslant c_1^\star + b_{\text{c}} - 1\}$, with $\text{card}({\mathcal C}_1) = b_r b_c$, where $r_1^\star$ and $c_1^\star$ are index parameters determined by
    \begin{equation}
    (r_1^\star,c_1^\star)=\arg\max_{r,c}\frac{1}{b_{\text{r}}b_{\text{c}}}\sum\limits_{i=r}^{r+b_{\text{r}}-1}\sum\limits_{j=c}^{c+b_{\text{c}}-1}|\underline{\widehat{\mathbf B}_1}[i,j]|. \notag
    \end{equation}
    The $b_r b_c$ candidate beam pairs in ${\mathcal C}_1$ are then trained and the beam pair $(i_1^{\star}, j_1^{\star})$ that leads to the highest measured output energy is identified as the analog beam pair for the first BS-UE link. 
    Similary, for the $u$-th ($u>1$) BS-UE link, we first set $\{\underline{\widehat{\mathbf B}_u}[:, j_1^\star],\dots,\underline{\widehat{\mathbf B}_u}[:, j_{u-1}^\star]\}$ to zeros for interference coordination purpose, and determine a subset of candidate mmWave beams ${\mathcal C}_u=\{(\mathbf{z}_{i}, \mathbf{w}_{j}) | r_u^\star \leqslant i \leqslant r_u^\star + b_{\text{r}} - 1, c_u^\star \leqslant j \leqslant c_u^\star + b_{\text{c}} - 1\}$, with $\text{card}({\mathcal C}_u) = b_r b_c$, and index parameters $r_u^{\star}$ and $c_u^{\star}$ determined by  
    \begin{equation}
    (r_u^\star,c_u^\star)=\arg\max_{r,c}\frac{1}{b_{\text{r}}b_{\text{c}}}\sum\limits_{i=r}^{r+b_{\text{r}}-1}\sum\limits_{j=c}^{c+b_{\text{c}}-1}|\underline{\widehat{\mathbf B}_u}[i,j]|. \notag
    \end{equation}
    We then identify the analog beam pair for the $u$-th BS-UE link $(i_u^{\star},j_u^{\star})$ by searching the beam pair from $\mathcal C_u$ that leads to the largest energy output through mmWave beam training.
    As for digital beamforming, LMMSE precoder is formed based on estimate of effective channel in digital domain as before. 
}
\end{itemize}

\par In addition, we also consider the integration of mmWave beamspace prediction network ${\mathbf G}(\cdot)$ with the above two baselines for comparison: 
\begin{itemize}
    \item {\bf ${\mathbf G}(\cdot)$+uncoordinated method}: For each BS-UE link, this method first uses ${\mathbf G}(\cdot)$ proposed to generate predicted mmWave beamspace representation $\widehat{\mathbf B}$, and then select the analog beam pair corresponding to maximum gain from $\widehat{\mathbf B}$ as the analog beams of this link. Following these, LMMSE precoder is formed based on the estimate of the effective channel in the digital domain, as in the previous methods.
    \item {\bf ${\mathbf G}(\cdot)$+coordinated method}: This method combines ${\mathbf G}(\cdot)$ for generating predicted mmWave beamspace representation $\widehat{\mathbf B}_u$, followed by a sequential mmWave analog beam selection process based on $\widehat{\mathbf B}_u$. Initially, the method selects the beam pair $(\mathbf{z}_{i_1^\star}, \mathbf{w}_{j_1^\star})$ as the analog precoder at the BS and combiner at the first UE, respectively, where $(i_1^\star, j_1^\star) = \arg\max_{i,j}|\widehat{{\mathbf B}}_1[i,j]|$. For subsequent iterations with $u\ge 2$, once $(i_1^\star, j_1^\star),\dots,(i_{u-1}^\star, j_{u-1}^\star)$ mmWave beam pairs have been selected for the previous $(u-1)$ BS-UE links, the rows ${{\widehat{\mathbf B}_u}[:, j_1^\star],\dots,{\widehat{\mathbf B}_u}[:, j_{u-1}^\star]}$ are zeroed out. Then, the beam pair with indices $(i_u^\star, j_u^\star) = \arg\max_{i,j}|\widehat{{\mathbf B}}_u[i,j]|$ is chosen for the $u$-th BS-UE link. 
    Regarding digital beamforming, LMMSE precoder is formed based on the estimate of the effective channel in the digital domain, as in the previous methods.
\end{itemize}

\par Table~\ref{Comparison of Baselines} highlights the differences between the baselines and the SA-MUHBF proposed.

\subsection{Performance Evaluation and Comparison}
\begin{table*}[t]
\renewcommand{\arraystretch}{2.2}
\caption{Comparison of Different Methods}
\label{Comparison of Baselines}
\centering
\setlength\tabcolsep{1.2mm}{
\begin{tabular}{c|c|c|c|c|c}
\toprule
Method& \makecell[c]{mmWave beamspace\\ representation prediction}  & Analog beam selection  & \makecell[c]{Digital \\beamforming} & \makecell[c]{Pilots \\at sub-6G}& \makecell[c]{Pilots \\at mmWave}\\ 
\midrule
Uncoordinated method & \XSolidBrush &\makecell[c]{uncoordinated beam selection \\based on sub-6G $\underline{\widehat{\mathbf B}}$} & LMMSE & \textcolor{black}{$K\underline{N_{\text{UE}}}$}&\textcolor{black}{$K b_r b_c +K$}\\ \hline
Coordinated method & \XSolidBrush & \makecell[c]{sequential beam selection \\based on sub-6G $\underline{\widehat{\mathbf B}}$} & LMMSE& \textcolor{black}{$K\underline{N_{\text{UE}}}$} & \textcolor{black}{$K b_r b_c +K$}\\ \hline
\makecell[c]{${\mathbf G}(\cdot)$ + \\uncoordinated method}& CNN-aided: ${\widehat{\mathbf B}}={\mathbf G}(\underline{\widehat{\mathbf B}})$&\makecell[c]{uncoordinated beam selection\\ based on predicted mmWave ${\widehat{\mathbf B}}$ }&LMMSE & \textcolor{black}{$K\underline{N_{\text{UE}}}$}& \textcolor{black}{$K$}\\ \hline
\makecell[c]{${\mathbf G}(\cdot)$ + \\coordinated method}& CNN-aided: ${\widehat{\mathbf B}}={\mathbf G}(\underline{\widehat{\mathbf B}})$&\makecell[c]{sequential beam selection \\based on predicted mmWave ${\widehat{\mathbf B}}$}&LMMSE & \textcolor{black}{$K\underline{N_{\text{UE}}}$}& \textcolor{black}{$K$}\\ \hline
\makecell[c]{Our proposed\\ SA-MUHBF}& CNN-aided: ${\widehat{\mathbf B}}={\mathbf G}(\underline{\widehat{\mathbf B}})$&\makecell[c]{GNN-aided: \\$\{\tilde{\mathbf L}_1,\dots,\tilde{\mathbf L}_K\}={\mathbf S}(\widehat{{\mathbf B}}_1\dots,\widehat{{\mathbf B}}_K)$}&LMMSE & \textcolor{black}{$K\underline{N_{\text{UE}}}$}& \textcolor{black}{$K$}\\
\bottomrule
\end{tabular}}
\end{table*}
We first show the effectiveness of mmWave beamspace prediction network ${\mathbf G}(\cdot)$ and demonstrate the overall performance superiority of SA-MUHBF over the aforementioned baselines.
 
\par
\begin{figure*}[t]  
    \centering  
    \subfloat[Instance 1]{  
        \includegraphics[width=0.32\textwidth]{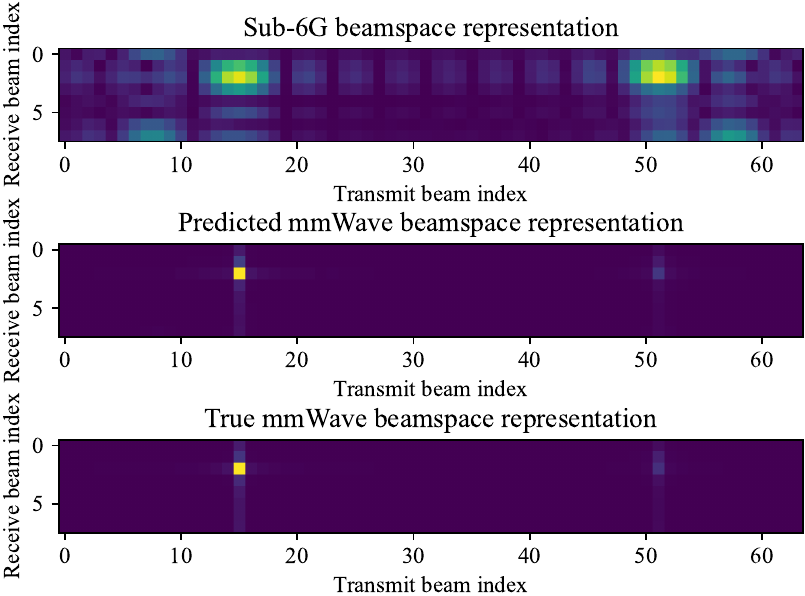}  
        \label{fig:sub1}  
    }  
    \subfloat[Instance 2]{  
        \includegraphics[width=0.32\textwidth]{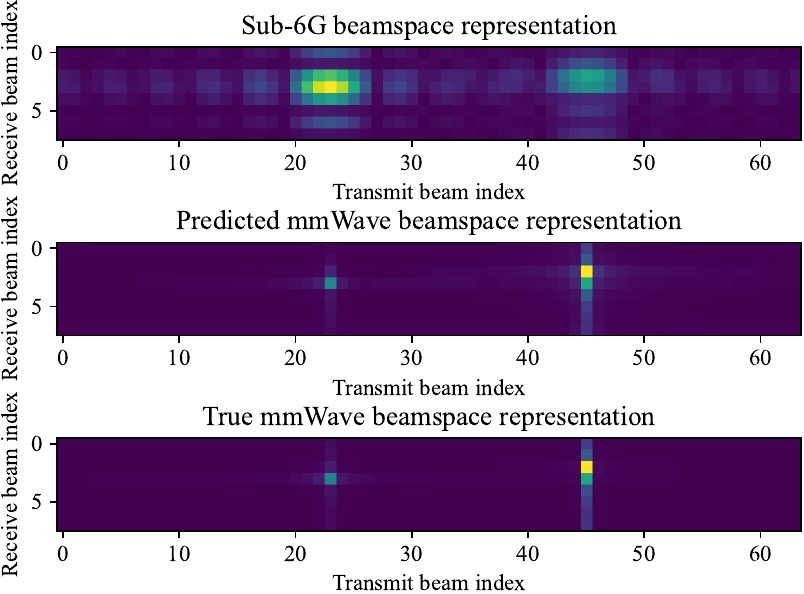}  
        \label{fig:sub2}  
    }  
    \subfloat[Instance 3]{  
        \includegraphics[width=0.32\textwidth]{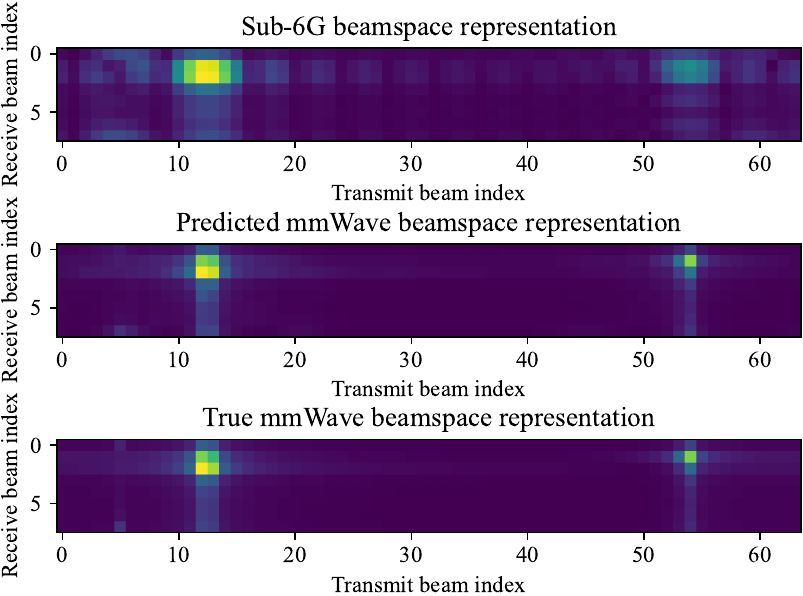}  
        \label{fig:sub3}  
    }  
    \caption{Comparison between sub-6G beamspace representation $\underline{\widehat{\mathbf B}}_u$, predicted mmWave beamspace representation $\widehat{\mathbf B}_u$, and true mmWave beamspace representation ${\mathbf B}_u$.}  
    \label{fig:total}  
\end{figure*}

 Specifically, in Fig.~\ref{fig:total}, we present the beamspace representations of the sub-6G channel estimate $\underline{\widehat{\mathbf B}}_u$, the predicted mmWave channel beamspace $\widehat{{\mathbf B}}_u$, and the actual mmWave channel beamspace ${\mathbf B}_u$ for three instances, with a sub-6G pilot signal power $\underline{P_{\text{s}}}=20~\text{dBm}$. 
It is evident that the prediction network ${\mathbf G}(\cdot)$ adeptly extracts spatial information from the sub-6G channel estimate, producing accurate predictions of ${\mathbf B}$, including both the angles and gains of dominant paths.
Moreover, we assess the NMSE between $\widehat{{\mathbf B}}_u$ and ${\mathbf B}_u$, as defined in~\eqref{l1}, averaged across all samples in the testing dataset. 
The resulting NMSE is merely 0.0894, further substantiating the effectiveness of ${\mathbf G}(\cdot)$.

\par
\begin{figure}[t]
\centering
\includegraphics[width=0.45\textwidth]{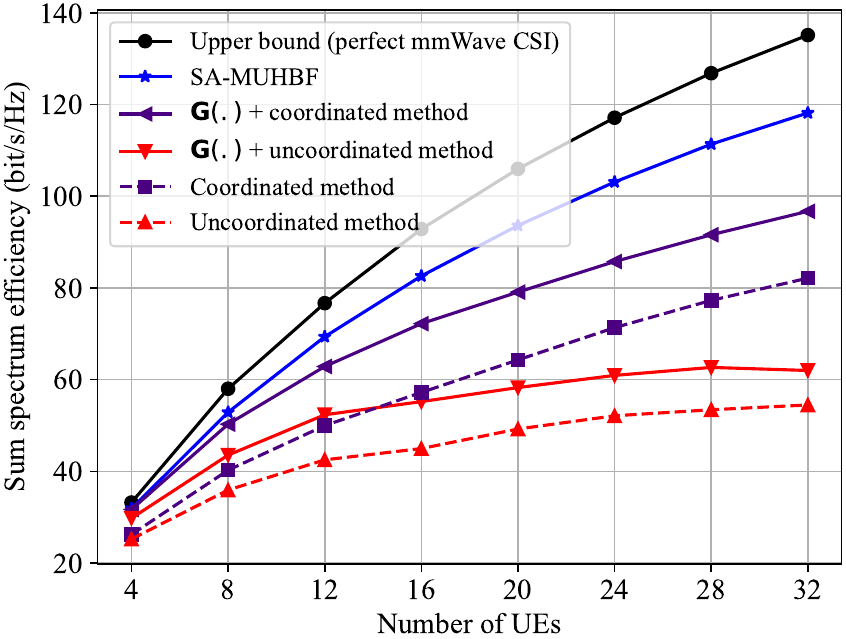}
\caption{Performance comparison between SA-MUHBF, four baselines, and an optimized hybrid beamforming with perfect mmWave CSI as the number of UEs increases.}
\label{SG effectiveness}
\end{figure}

 After fine-tuning ${\mathbf G}(\cdot)$, we next evaluate the performance of SA-MUHBF and compare against the baselines.  
Fig.~\ref{SG effectiveness} plots the sum spectrum efficiency of different methods under different number of UEs $K$.
It can be seen that ``${\mathbf G}(\cdot)$+uncoordinated method" and ``${\mathbf G}(\cdot)$+coordinated method" outperform their respective counterparts, ``uncoordinated method" and ``coordinated method". 
This underscores the importance of translating sub-6G channel estimates into mmWave beam space representation for effective mmWave beam selection. 
Our proposed SA-MUHBF consistently outperforms all baselines based on sub-6G information. 
Specifically, we observe performance gains of 0.5\% at 4 UEs, 5.2\% at 8 UEs, 14.4\% at 16 UEs, and 22.2\% at 32 UEs over the best-performing baseline. 
Notably, this improvement becomes more pronounced with increasing $K$, suggesting that the GNN-based network ${\mathbf S}(\cdot)$ in SA-MUHBF is highly effective in optimizing analog beam selection through learned graph convolutions, as opposed to the lack of interference coordination or heuristic interference coordination present in the baseline methods.
\textcolor{black}{Furthermore, we compare the performance of SA-MUHBF with an optimized hybrid beamforming scheme that assumes perfect mmWave CSI, using alternating minimization and manifold optimization as in \cite{10187715}. 
The sum spectral efficiency of this optimized scheme serves as an upper bound for all the methods considered. 
The results show that SA-MUHBF performs close to this upper bound, achieving 91.15\% efficiency with 8 UEs, 88.97\% with 16 UEs, and 87.44\% with 32 UEs. 
These findings further validate the effectiveness of using sub-6G CSI to predict mmWave CSI and perform beamforming within the SA-MUHBF framework.}

\begin{figure}[t]
    \centering
    \subfloat[]{  
            \includegraphics[width=0.45\textwidth]{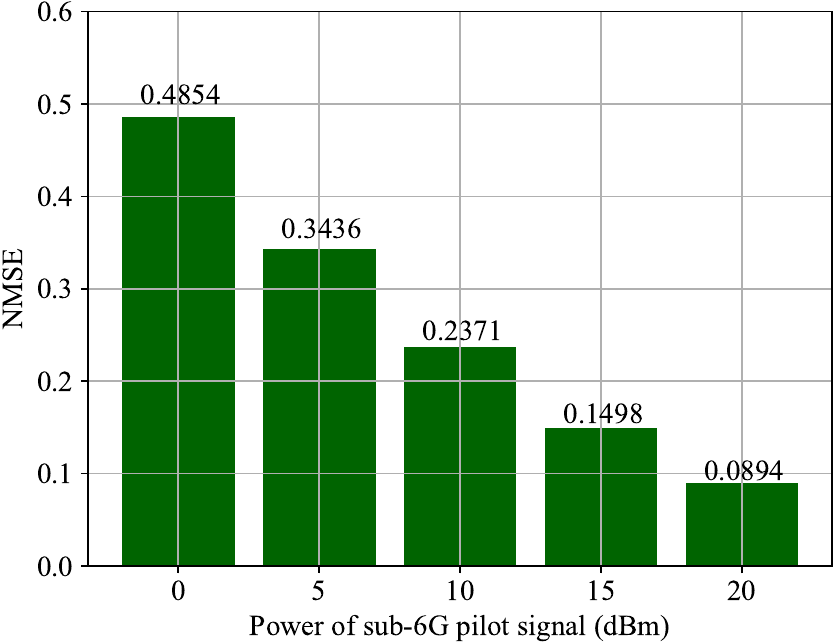}  
            \label{sub-6G power:sub1}  
        }\\
        \subfloat[]{  
            \includegraphics[width=0.45\textwidth]{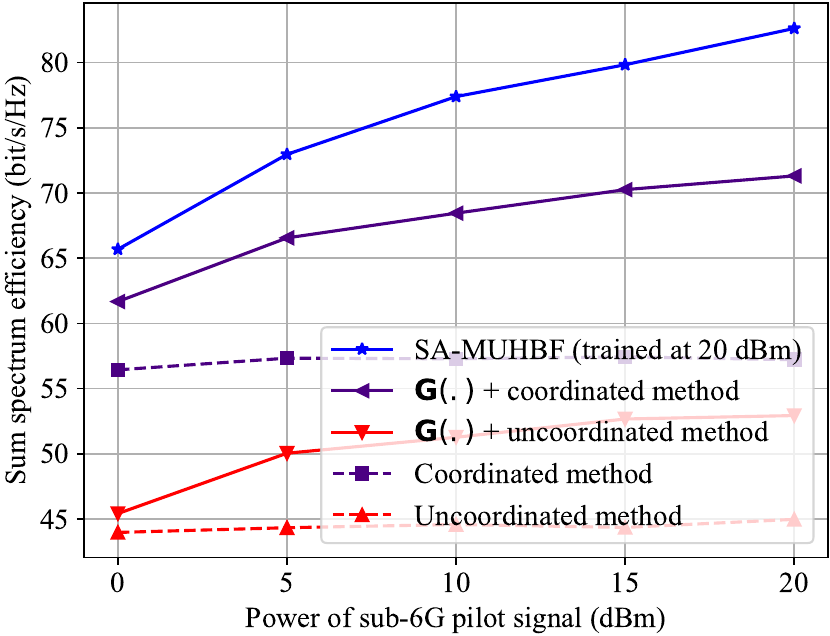}  
            \label{sub-6G power:sub2}  
        } 
    \caption{Performance comparison as $\underline{P_{\text{s}}}$ varies, when the number of UEs $K$ is fixed at 16, and the number of sub-6G antennas is fixed at $(4,16)$: (a) the prediction NMSE of ${\mathbf G}(\cdot)$ under different $\underline{P_{\text{s}}}$ values, (b) the sum spectrum efficiency achieved by different methods.}
    \label{sub-6G power}
    \end{figure}

\par Next, we adopt the ${\mathbf G}(\cdot)$ and ${\mathbf S}(\cdot)$ models, trained with pilot signal power $\underline{P_s}=20~\text{dBm}$ as above, to assess the performance of SA-MUHBF across various $\underline{P_{\text{s}}}$ values that directly impact the quality of sub-6G channel estimate. 
Fig.~\ref{sub-6G power:sub1} shows the NMSE of ${\mathbf G}(\cdot)$ versus $\underline{P_s}$ for SA-MUHBF, while Fig.~\ref{sub-6G power:sub2} illustrates the achievable spectrum efficiency versus $\underline{P_s}$ under different methods with $K=16$ UEs. 
It is evident that SA-MUHBF consistently outperforms all considered baselines and maintains significant gain over the best-performing baseline. 
These results confirm the robustness of SA-MUHBF across varying qualities of sub-6G channel estimates for optimizing mmWave beamforming.  

\par
\begin{figure}[t]
\centering
\subfloat[]{  
        \includegraphics[width=0.45\textwidth]{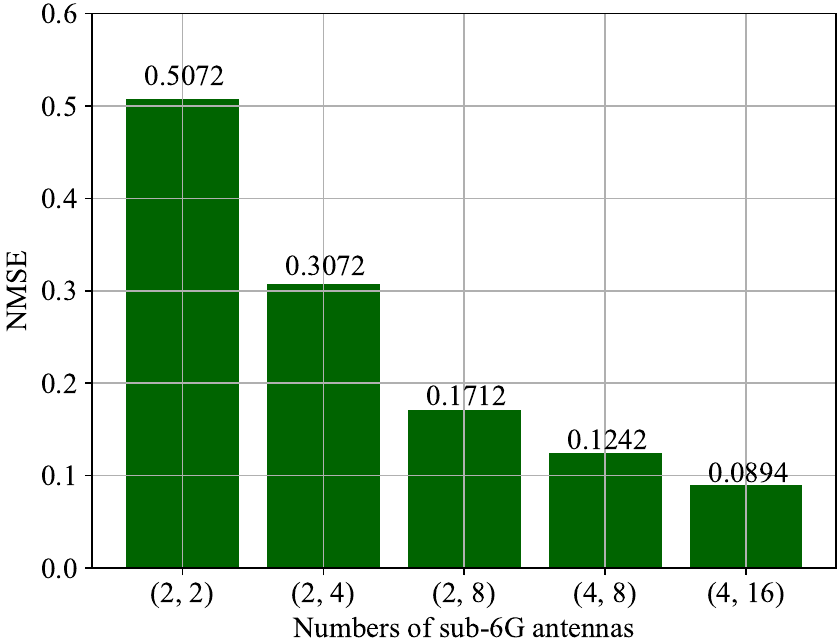}  
        \label{sub-6G antenna:sub1}  
    }\\
    \subfloat[]{  
        \includegraphics[width=0.45\textwidth]{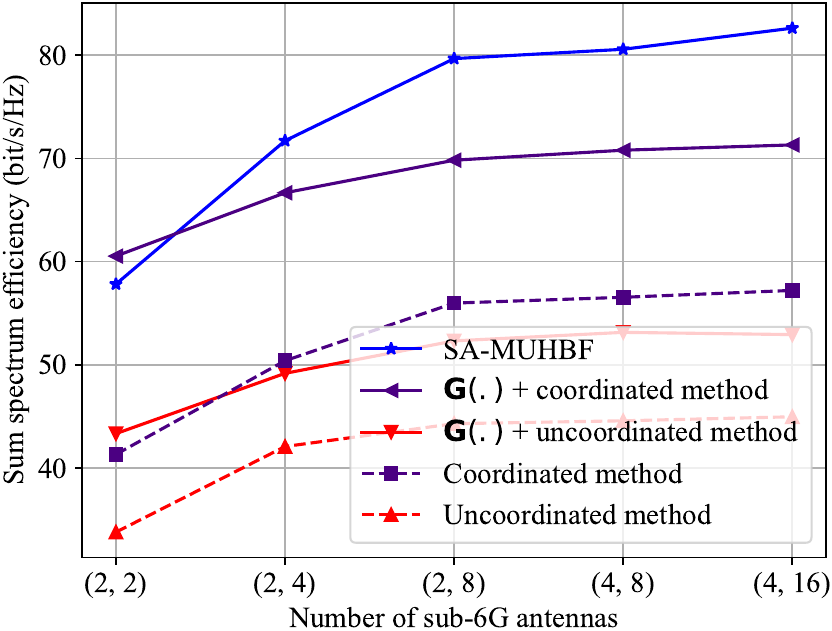}  
        \label{sub-6G antenna:sub2}  
    } 
\caption{Performance comparison as $(\underline{N_{\text{UE}}},~\underline{N_{\text{BS}}})$ varies, when the number of UEs $K$ is fixed at 16, and the power of sub-6G pilot signal power is fixed at 20 dBm: (a) the prediction NMSE of ${\mathbf G}(\cdot)$ under different sub-6G antenna setups, (b) the sum spectrum efficiency achieved by different methods.}
\label{sub-6G antenna}
\end{figure}

 Furthermore, we investigate the performance of SA-MUHBF under various sub-6G antenna configurations $(\underline{N_{\text{UE}}},\underline{N_{\text{BS}}})$. 
It is important to note that different antenna setups impact the angular resolution of the sub-6G channel estimate and the dimension of the sub-6G beamspace representation. 
Hence, we train separate ${\mathbf G}(\cdot)$ and ${\mathbf S}(\cdot)$ models for each considered configuration. 
Fig.~\ref{sub-6G antenna} displays the results for $(\underline{N_{\text{UE}}},\underline{N_{\text{BS}}})$ combinations including $(2,~2)$, $(2,~4)$, $(2,~8)$, $(4,~8)$, and $(4,~16)$. 
With small values for $\underline{N_{\text{UE}}}$ or $\underline{N_{\text{BS}}}$, the angular resolution in the sub-6G channel estimate is limited, making it relatively challenging to extract mmWave spatial information, as indicated by the high NMSE of ${\mathbf G}(\cdot)$ shown in Fig.~\ref{sub-6G antenna:sub1} when $(\underline{N_{\text{UE}}},\underline{N_{\text{BS}}}) = (2,2)$. 
However, with slight increases in $\underline{N_{\text{UE}}}$ or $\underline{N_{\text{BS}}}$, the NMSE of mmWave beamspace prediction decreases. 
This improvement is attributed to the remarkable capability of CNNs in representation learning within the ${\mathbf G}(\cdot)$ model. Fig.~\ref{sub-6G antenna:sub2} illustrates the achieved spectrum efficiency under different $(\underline{N_{\text{UE}}},\underline{N_{\text{BS}}})$ settings. 
The superiority of SA-MUHBF is again confirmed in most settings, despite a slight loss when $(\underline{N_{\text{UE}}},\underline{N_{\text{BS}}}) = (2,2)$ due to ineffective mmWave beamspace prediction in such case. 

\subsection{Generalization Capability of SA-MUHBF in Unseen Scenarios}
\begin{figure}[t]
\centering
\subfloat[]{  
        \includegraphics[width=0.45\textwidth]{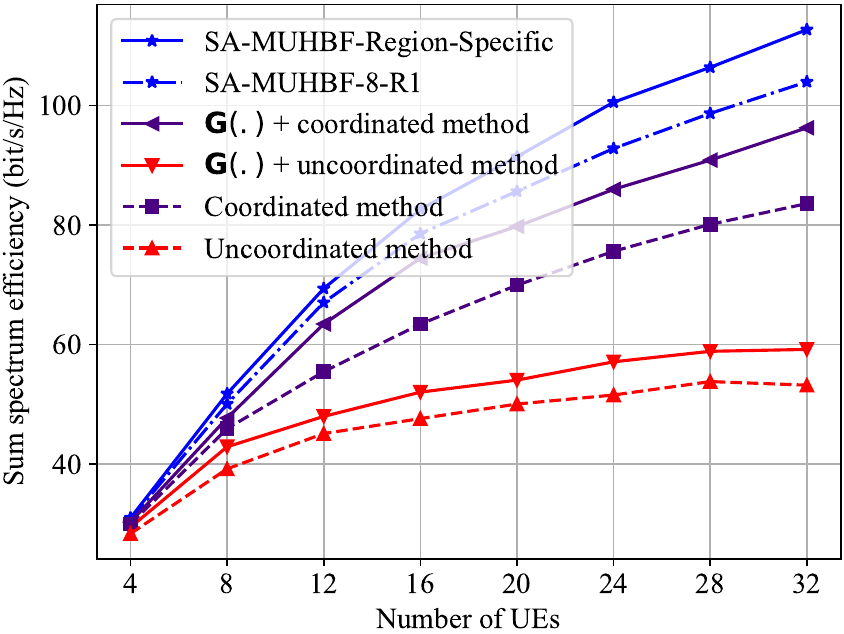}  
        \label{Generalization performance:sub1}  
    }\\
    \subfloat[]{  
        \includegraphics[width=0.45\textwidth]{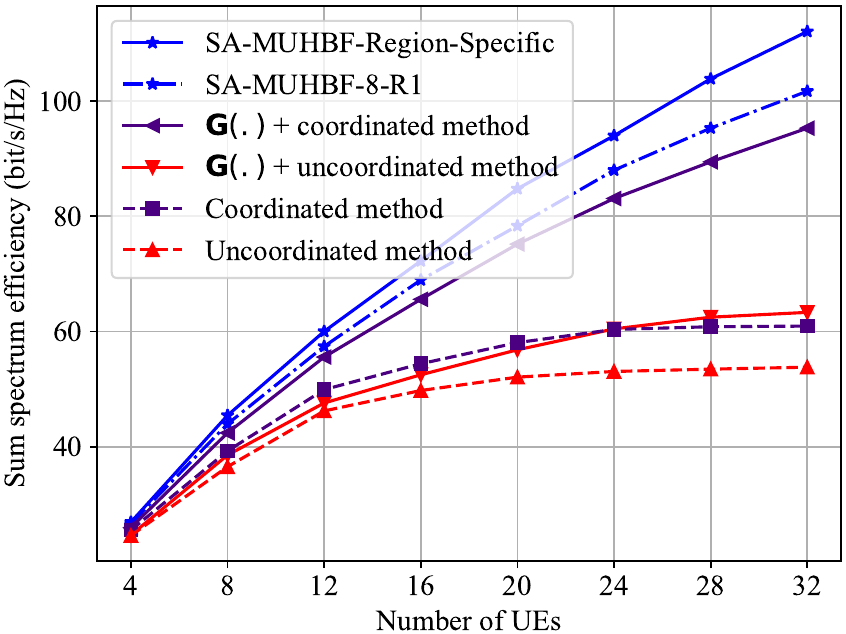}  
        \label{Generalization performance:sub2}  
    } 
\caption{Performance of SA-MUHBF for unseen regions: (a) evaluation with testing samples generated from BS 3 and Region 2, (b) evaluation with testing samples generated from BS 16 and Region 3.}
\label{Generalization performance}
\end{figure}

We now showcase the generalization capability of SA-MUHBF by utilizing samples from Region 2 and Region 3 (see Fig.~\ref{O1 top} for illustration) as testing samples. 
Notably, for the generalization test, SA-MUHBF is trained for the 8 UEs setup in Region 1, denoted as ``SA-MUHBF-8-R1'', \textcolor{black}{and SA-MUHBF models specifically trained for each setup in regions 2 and 3, are denoted by ``SA-MUHBF-Region-Specific''}. 
The prediction NMSEs of ${\mathbf G}(\cdot)$ for SA-MUHBF-8-R1 on the samples from Region 2 and Region 3 are $0.1843$ and $0.1612$, respectively.
\textcolor{black}{Meanwhile, the specifically trained ${\mathbf G}(\cdot)$ achieves lower NMSEs of 0.0946 and 0.0857 for regions 2 and 3, respectively.}
Furthermore, Fig.~\ref{Generalization performance} illustrates the sum spectrum efficiency of SA-MUHBF, along with the aforementioned baseline algorithms, on the testing samples from regions 2 and 3. 
As depicted, SA-MUHBF-8-R1 surpasses all baseline algorithms, demonstrating gains of 1.6\% at 4 UEs, 4.9\% at 8 UEs, 5.5\% at 8 UEs, and 8.0\% at 32 UEs on the testing samples from Region 2, and gains of 2.5\% at 4 UEs, 3.7\% at 8 UEs, 5.1\% at 16 UEs, and 6.7\% at 32 UEs on the testing samples from Region 3, over the best performing baseline. 
\textcolor{black}{Additionally, the SA-MUHBF-8-R1 model achieves 99.35\% and 92.25\% of the performance of region-specific SA-MUHBF models for Region 2 when the number of UEs is 4 and 32, respectively.
For Region 3, it achieves 98.89\% and 90.76\% under the same UE conditions.
When the number of UEs is small, despite regional differences, the potential interference between BS-UE links is relatively manageable, allowing the SA-MUHBF-8-R1 model to generalize well to unseen scenarios.
With a larger number of UEs, increased regional differences and more complex interference reduce the performance of the SA-MUHBF-8-R1 model compared to region-specific models, though it still demonstrates significant generalization capabilities.}

\subsection{Algorithm Pilot Overhead and Execution Complexity Analysis}
\subsubsection{Pilot Overhead Analysis}
\begin{figure}[t]
\centering
\includegraphics[width=0.45\textwidth]{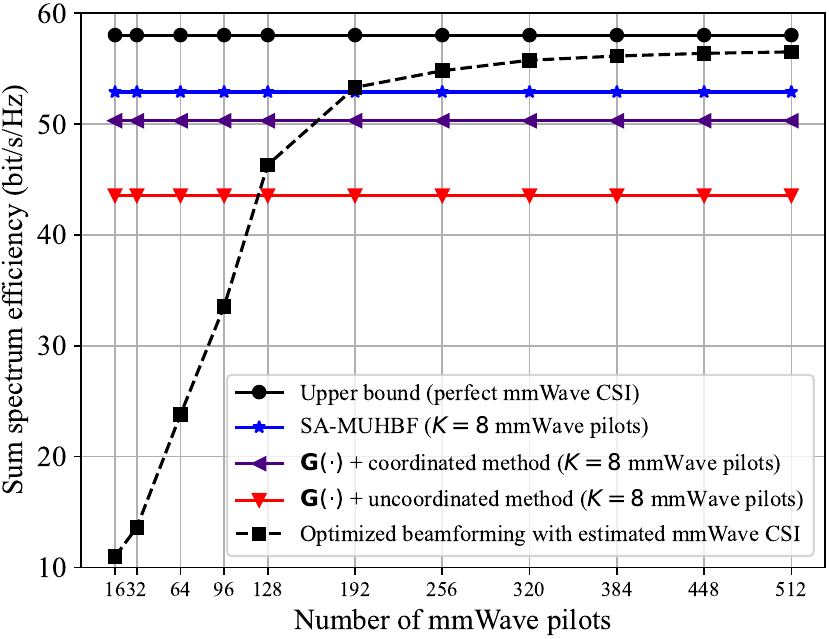}
\caption{Performance comparison between SA-MUHBF, ${\mathbf G}(\cdot)$ + uncoordinated method, ${\mathbf G}(\cdot)$ + coordinated  method, optimized  beamforming with estimated mmWave CSI and an upper bound with perfect mmWave CSI.}
\label{fig:SA_MUHBF_baselines_pilot_based}
\end{figure}

Here, we quantify the required pilot overhead at sub-6G and mmWave frequencies of our proposed SA-MUHBF and the baselines considered, see Table~\ref{Comparison of Baselines}.
Moreover, to illustrate the overhead reduction compared to estimated mmWave CSI, we also consider a scheme that begins with compressive-sensing-based mmWave channel estimation using in-band pilot measurements \cite{alkhateeb2015compressed}. 
Following this, optimized hybrid beamforming is designed using the estimated mmWave CSI through alternating minimization and manifold optimization \cite{10187715}. 
We evaluate the achievable sum spectral efficiency of this scheme under various pilot measurement configurations in Fig.~\ref{fig:SA_MUHBF_baselines_pilot_based} and demonstrate that the sub-6G CSI-aided methods can achieve the same performance with reduced pilot overhead.
Specifically, to match the performance of SA-MUHBF, ``${\mathbf G}(\cdot)$ + coordinated method'', and ``${\mathbf G}(\cdot)$ + uncoordinated method'', the in-band mmWave channel estimation method needs at least 192, 160, and 128 mmWave pilots, respectively.
In constrast, the sub-6G CSI driven methods considered require only $K\underline{N_{\text{UE}}}=32$ sub-6G pilots and $K=8$ mmWave pilots.
The pilot reduction is attributed to the spatial congruence between sub-6G and mmWave channels, which reduces the need for extensive mmWave pilot measurements and facilitates mmWave analog beamforming. 
Additionally, the fully digital architecture in the sub-6G system enables efficient acquisition of sub-6G channel estimates.

\subsubsection{Execution Complexity Analysis}
We now analyze the execution complexity of SA-MUHBF and the considered baselines.
For SA-MUHBF, as depicted in Fig.~\ref{diagram}, we have: (\textit{i}) its beamspace prediction ${\mathbf G}(\cdot)$ has a complexity on the order of ${\mathcal O}(KT_{\text{G}} N_{\text{z}} N_{\text{w}})$, where $K$ is the number of beams, $T_{\text{G}}$ is the number of layers in ${\mathbf G}(\cdot)$;
(\textit{ii}) the analog beam selection ${\mathbf S}(\cdot)$ has a complexity on the order of ${\mathcal O}(K^2N_{\text{z}}N_{\text{w}}T)$, where $T$ is the number of layers in ${\mathbf S}(\cdot)$;
(\textit{iii}) the linear MMSE precoder has a complexity of ${\mathcal O}(N_{\text{RF}}^3)$.
Consequently, the overall complexity of SA-MUHBF is on the order of ${\mathcal O}((K^2T+KT_{\text{G}})N_{\text{z}}N_{\text{w}}+N_{\text{RF}}^3)$.
As for the considered baselines, both ``uncoordinated method'' and ``coordinated method'' have a complexity of ${\mathcal O}(K N_{\text{z}} N_{\text{w}}+N_{\text{RF}}^3)$, while both ``${\mathbf G}(\cdot)$ + uncoordinated method'' and ``${\mathbf G}(\cdot)$ + coordinated method'' have a complexity of ${\mathcal O}((KT_{\text{G}}+K)N_{\text{z}}N_{\text{w}}+N_{\text{RF}}^3)$.
It is noted that although SA-MUHBF has a slightly higher complexity order than the baselines, its actual execution time could be lower due to the use of graphics processing unit acceleration for the CNN and GNN implementations. 
For example, in a scenario with 12 UEs, SA-MUHBF is 5 times faster than the ``uncoordinated method'' in our experiment.

\section{Conclusion}
\label{Conclusion}
In this work, we have exploited the spatial congruence between sub-6G channel and mmWave channel to facilitate the multi-user mmWave hybrid beamforming.
A deep learning based framework, named SA-MUHBF, has been developed to achieve this end. SA-MUHBF utilizes a convolutional neural network to predict mmWave beamspace from sub-6G channel estimate, 
followed by a customized multi-layer graph neural network for analog beam selection and a LMMSE precoder for digital beamforming.  
Extensive numerical experiments validate the effectiveness of SA-MUHBF, demonstrating its superiority over several state-of-the-art benchmarks. 
Notably, SA-MUHBF requires reduced pilot overhead at mmWave and exhibits robust performance across various system configurations and unseen scenarios.

\bibliographystyle{IEEEtran}
\bibliography{IEEEabrv,reference_simplified}

\providecommand{\noopsort}[1]{}
\begin{thebibliography}{10}
\providecommand{\url}[1]{#1}
\csname url@samestyle\endcsname
\providecommand{\newblock}{\relax}
\providecommand{\bibinfo}[2]{#2}
\providecommand{\BIBentrySTDinterwordspacing}{\spaceskip=0pt\relax}
\providecommand{\BIBentryALTinterwordstretchfactor}{4}
\providecommand{\BIBentryALTinterwordspacing}{\spaceskip=\fontdimen2\font plus
\BIBentryALTinterwordstretchfactor\fontdimen3\font minus
  \fontdimen4\font\relax}
\providecommand{\BIBforeignlanguage}[2]{{%
\expandafter\ifx\csname l@#1\endcsname\relax
\typeout{** WARNING: IEEEtran.bst: No hyphenation pattern has been}%
\typeout{** loaded for the language `#1'. Using the pattern for}%
\typeout{** the default language instead.}%
\else
\language=\csname l@#1\endcsname
\fi
#2}}
\providecommand{\BIBdecl}{\relax}
\BIBdecl

\bibitem{8454665}
C.~Liu, M.~Li, S.~V. Hanly, P.~Whiting, and I.~B. Collings, ``Millimeter-wave
  small cells: Base station discovery, beam alignment, and system design
  challenges,'' \emph{{IEEE} Wireless Commun.}, vol.~25, no.~4, pp. 40--46,
  2018.

\bibitem{Alkhateeb2014}
A.~Alkhateeb, O.~El~Ayach, G.~Leus, and R.~W. Heath, ``Channel estimation and
  hybrid precoding for millimeter wave cellular systems,'' \emph{{IEEE} J. Sel.
  Topics Signal Process}, vol.~8, no.~5, pp. 831--846, 2014.

\bibitem{heath2016overview}
R.~W. Heath, N.~González-Prelcic, S.~Rangan, W.~Roh, and A.~M. Sayeed, ``An
  overview of signal processing techniques for millimeter wave mimo systems,''
  \emph{{IEEE} Trans. Signal Process.}, vol.~10, no.~3, pp. 436--453, 2016.

\bibitem{sohrabiHybridDigitalAnalog2016}
F.~Sohrabi and W.~Yu, ``Hybrid digital and analog beamforming design for
  large-scale antenna arrays,'' \emph{{IEEE} J. Sel. Topics Signal Process},
  vol.~10, no.~3, pp. 501--513, Apr. 2016.

\bibitem{alkhateeb2016frequency}
A.~Alkhateeb and R.~W. Heath, ``Frequency selective hybrid precoding for
  limited feedback millimeter wave systems,'' \emph{{IEEE} Trans. Wireless
  Commun.}, vol.~64, no.~5, pp. 1801--1818, 2016.

\bibitem{huTwoTimescaleEndtoEndLearning2022}
Q.~Hu, Y.~Cai, K.~Kang, G.~Yu, J.~Hoydis, and Y.~C. Eldar, ``Two-timescale
  end-to-end learning for channel acquisition and hybrid precoding,''
  \emph{{IEEE} J. Select. Areas Commun.}, vol.~40, no.~1, pp. 163--181, Jan.
  2022.

\bibitem{alkhateeb2015limited}
A.~Alkhateeb, G.~Leus, and R.~W. Heath, ``Limited feedback hybrid precoding for
  multi-user millimeter wave systems,'' \emph{IEEE Trans. Wireless Commun.},
  vol.~14, no.~11, pp. 6481--6494, 2015.

\bibitem{9045972}
S.~S. Nair and S.~Bhashyam, ``Hybrid beamforming in {MU-MIMO} using partial
  interfering beam feedback,'' \emph{{IEEE} Commun. Lett.}, vol.~24, no.~7, pp.
  1548--1552, 2020.

\bibitem{10187715}
M.~Hui, X.~Zhao, T.~Lin, and Y.~Zhu, ``Hybrid beamforming for utility
  maximization in multiuser broadband millimeter wave systems,'' \emph{{IEEE}
  Trans. Veh. Technol.}, vol.~72, no.~12, pp. 16\,042--16\,057, 2023.

\bibitem{elbir2019hybrid}
A.~M. Elbir and A.~K. Papazafeiropoulos, ``Hybrid precoding for multiuser
  millimeter wave massive mimo systems: A deep learning approach,''
  \emph{{IEEE} Trans. Veh. Technol.}, vol.~69, no.~1, pp. 552--563, 2019.

\bibitem{jinModelDrivenDeepLearning2023}
W.~Jin, J.~Zhang, C.-K. Wen, and S.~Jin, ``Model-driven deep learning for
  hybrid precoding in millimeter wave {MU-MIMO} system,'' \emph{{IEEE} Trans.
  Wireless Commun.}, vol.~71, no.~10, pp. 5862--5876, 2023.

\bibitem{peter2016measurement}
M.~Peter, K.~Sakaguchi, S.~Jaeckel, S.~Wu, M.~Nekovee, J.~Medbo, K.~Haneda,
  S.~Nguyen, R.~Naderpour, J.~Vehmas \emph{et~al.}, ``Measurement campaigns and
  initial channel models for preferred suitable frequency ranges,''
  \emph{Deliverable D2}, vol.~1, p. 160, 2016.

\bibitem{samimi20163}
M.~K. Samimi and T.~S. Rappaport, ``{3-D} millimeter-wave statistical channel
  model for 5g wireless system design,'' \emph{{IEEE} Trans. Microwave Theory
  Tech.}, vol.~64, no.~7, pp. 2207--2225, 2016.

\bibitem{9768944}
D.~Dupleich, N.~Han, A.~Ebert, R.~Müller, S.~Ludwig, A.~Artemenko,
  J.~Eichinger, T.~Geiss, G.~Del~Galdo, and R.~Thomä, ``From sub-6 {GHz} to
  mm-wave: Simultaneous multi-band characterization of propagation from
  measurements in industry scenarios,'' in \emph{Eur. Conf. Antennas Propag.},
  2022, pp. 1--5.

\bibitem{10058899}
P.~Kyösti, P.~Zhang, A.~Pärssinen, K.~Haneda, P.~Koivumäki, and W.~Fan, ``On
  the feasibility of out-of-band spatial channel information for
  millimeter-wave beam search,'' \emph{{IEEE} Trans. Antennas Propagat.},
  vol.~71, no.~5, pp. 4433--4443, 2023.

\bibitem{nitsche2015steering}
T.~Nitsche, A.~B. Flores, E.~W. Knightly, and J.~Widmer, ``Steering with eyes
  closed: mm-wave beam steering without in-band measurement,'' in \emph{Proc.
  IEEE INFOCOM}, 2015, pp. 2416--2424.

\bibitem{7888146}
A.~Ali, N.~González-Prelcic, and R.~W. Heath, ``Estimating millimeter wave
  channels using out-of-band measurements,'' in \emph{Inf. Theory Appl.
  Workshop}, 2016, pp. 1--6.

\bibitem{8792393}
A.~Ali, N.~González-Prelcic, and R.~W. Heath, ``Spatial covariance estimation for millimeter wave hybrid systems
  using out-of-band information,'' \emph{{IEEE} Trans. Wireless Commun.},
  vol.~18, no.~12, pp. 5471--5485, 2019.

\bibitem{ali2017millimeter}
A.~Ali, N.~Gonz{\'a}lez-Prelcic, and R.~W. Heath, ``Millimeter wave
  beam-selection using out-of-band spatial information,'' \emph{IEEE Trans.
  Wireless Commun.}, vol.~17, no.~2, pp. 1038--1052, 2017.

\bibitem{alrabeiah2020deep}
M.~Alrabeiah and A.~Alkhateeb, ``Deep learning for mmwave beam and blockage
  prediction using sub-6 {GHz} channels,'' \emph{{IEEE} Trans. Wireless
  Commun.}, vol.~68, no.~9, pp. 5504--5518, 2020.

\bibitem{10292615}
K.~Vuckovic, M.~B. Mashhadi, F.~Hejazi, N.~Rahnavard, and A.~Alkhateeb,
  ``Paramount: Toward generalizable deep learning for mmwave beam selection
  using sub-6 {GHz} channel measurements,'' \emph{{IEEE} Trans. Wireless
  Commun.}, vol.~23, no.~5, pp. 5187--5202, 2024.

\bibitem{10229493}
J.~Liu, X.~Li, T.~Fan, S.~Lv, and M.~Shi, ``Multimodal fusion assisted mmwave
  beam training in dual-model networks,'' \emph{{IEEE} Trans. Veh. Technol.},
  vol.~73, no.~1, pp. 995--1011, 2024.

\bibitem{dengwcnc2024}
W.~Deng, M.~Li, Y.~Liu, M.-M. Zhao, and M.~Lei, ``Enhancing mmwave beam
  prediction through deep learning with sub-6 {GHz} channel estimate,'' in
  \emph{{IEEE} Wireless Commun. Networking Conf.}, 2024, pp. 1--6.

\bibitem{ma2021deep}
K.~Ma, D.~He, H.~Sun, and Z.~Wang, ``Deep learning assisted mmwave beam
  prediction with prior low-frequency information,'' in \emph{IEEE Int. Conf.
  Commun.}\hskip 1em plus 0.5em minus 0.4em\relax IEEE, 2021, pp. 1--6.

\bibitem{maschiettiCoordinatedBeamSelection2019}
F.~Maschietti, D.~Gesbert, and P.~{\noopsort{kerret}}{de Kerret}, ``Coordinated
  beam selection in millimeter wave multi-user {MIMO} using out-of-band
  information,'' in \emph{IEEE Int. Conf. Commun.}, 2019, pp. 1--6.

\bibitem{liHybridPrecodingUsing2020}
Z.~Li, C.~Zhang, I.-T. Lu, and X.~Jia, ``Hybrid precoding using out-of-band
  spatial information for multi-user multi-rf-chain millimeter wave systems,''
  \emph{IEEE Access}, vol.~8, pp. 50\,872--50\,883, 2020.

\bibitem{liuCollaborativeManagementResource2023}
J.~Liu, X.~Li, T.~Fan, S.~Lv, and M.~Shi, ``Collaborative {{Management}} of
  {{Resource Allocation}} and {{Precoding}} for {{Dual-mode Networks}},''
  \emph{{IEEE} Trans. Veh. Technol.}, pp. 1--15, 2023.

\bibitem{he2021overview}
S.~He, S.~Xiong, Y.~Ou, J.~Zhang, J.~Wang, Y.~Huang, and Y.~Zhang, ``An
  overview on the application of graph neural networks in wireless networks,''
  \emph{IEEE open j. Commun. Soc.}, 2021.

\bibitem{shen2020graph}
Y.~Shen, Y.~Shi, J.~Zhang, and K.~B. Letaief, ``Graph neural networks for
  scalable radio resource management: Architecture design and theoretical
  analysis,'' \emph{{IEEE} J. Select. Areas Commun.}, vol.~39, no.~1, pp.
  101--115, 2020.

\bibitem{lee2020graph}
M.~Lee, G.~Yu, and G.~Y. Li, ``Graph embedding-based wireless link scheduling
  with few training samples,'' \emph{IEEE Trans. Wireless Commun.}, vol.~20,
  no.~4, pp. 2282--2294, 2020.

\bibitem{he2022gblinks}
S.~He, S.~Xiong, W.~Zhang, Y.~Yang, J.~Ren, and Y.~Huang, ``Gblinks:
  {GNN}-based beam selection and link activation for ultra-dense {D2D} mmwave
  networks,'' \emph{{IEEE} Trans. Wireless Commun.}, 2022.

\bibitem{10184114}
W.~Deng, Y.~Liu, M.~Li, and M.~Lei, ``{GNN}-aided user association and beam
  selection for mmwave-integrated heterogeneous networks,'' \emph{{IEEE}
  Wireless Commun. Lett.}, vol.~12, no.~11, pp. 1836--1840, 2023.

\bibitem{Alkhateeb2019}
A.~Alkhateeb, ``{DeepMIMO}: A generic deep learning dataset for millimeter wave
  and massive {MIMO} applications,'' in \emph{Inf. Theory Appl. Workshop}, San
  Diego, CA, Feb 2019, pp. 1--8.

\bibitem{Remcom}
Remcom, ``{Wireless InSite},'' \url{http://www.remcom.com/wireless-insite}.

\bibitem{alkhateeb2015compressed}
A.~Alkhateeb, G.~Leus, and R.~W. Heath, ``Compressed sensing based multi-user
  millimeter wave systems: How many measurements are needed?'' in \emph{IEEE
  Int. Conf. Acoust. Speech Signal Process}.\hskip 1em plus 0.5em minus
  0.4em\relax IEEE, 2015, pp. 2909--2913.

\bibitem{6472238}
Y.~Bengio, A.~Courville, and P.~Vincent, ``Representation learning: A review
  and new perspectives,'' \emph{{IEEE} Trans. Pattern Anal. Machine Intell.},
  vol.~35, no.~8, pp. 1798--1828, 2013.

\bibitem{shiMaskedLabelPrediction2021a}
Y.~Shi, Z.~Huang, S.~Feng, H.~Zhong, W.~Wang, and Y.~Sun, ``Masked label
  prediction: {Unified} message passing model for semi-supervised
  classification,'' in \emph{Int. Joint Conf. Artif. Intell.}, Aug. 2021, pp.
  1548--1554.

\bibitem{nguyen2014mmse}
D.~H. Nguyen and T.~Le-Ngoc, ``{MMSE} precoding for multiuser {MISO} downlink
  transmission with non-homogeneous user snr conditions,'' \emph{Eurasip. J.
  Adv. Sign. Process.}, vol. 2014, no.~1, pp. 1--12, 2014.

\bibitem{noauthor_reducelronplateau_nodate}
\BIBentryALTinterwordspacing
``{ReduceLROnPlateau} — {PyTorch} 2.1 documentation.'' [Online]. Available:
  \url{https://pytorch.org/docs/stable/generated/torch.optim.lr_scheduler.ReduceLROnPlateau.html}
\BIBentrySTDinterwordspacing

\end{thebibliography}
\vspace{-10mm}
\begin{IEEEbiography}
[{\includegraphics[width=1in,height=1.25in,clip,keepaspectratio]{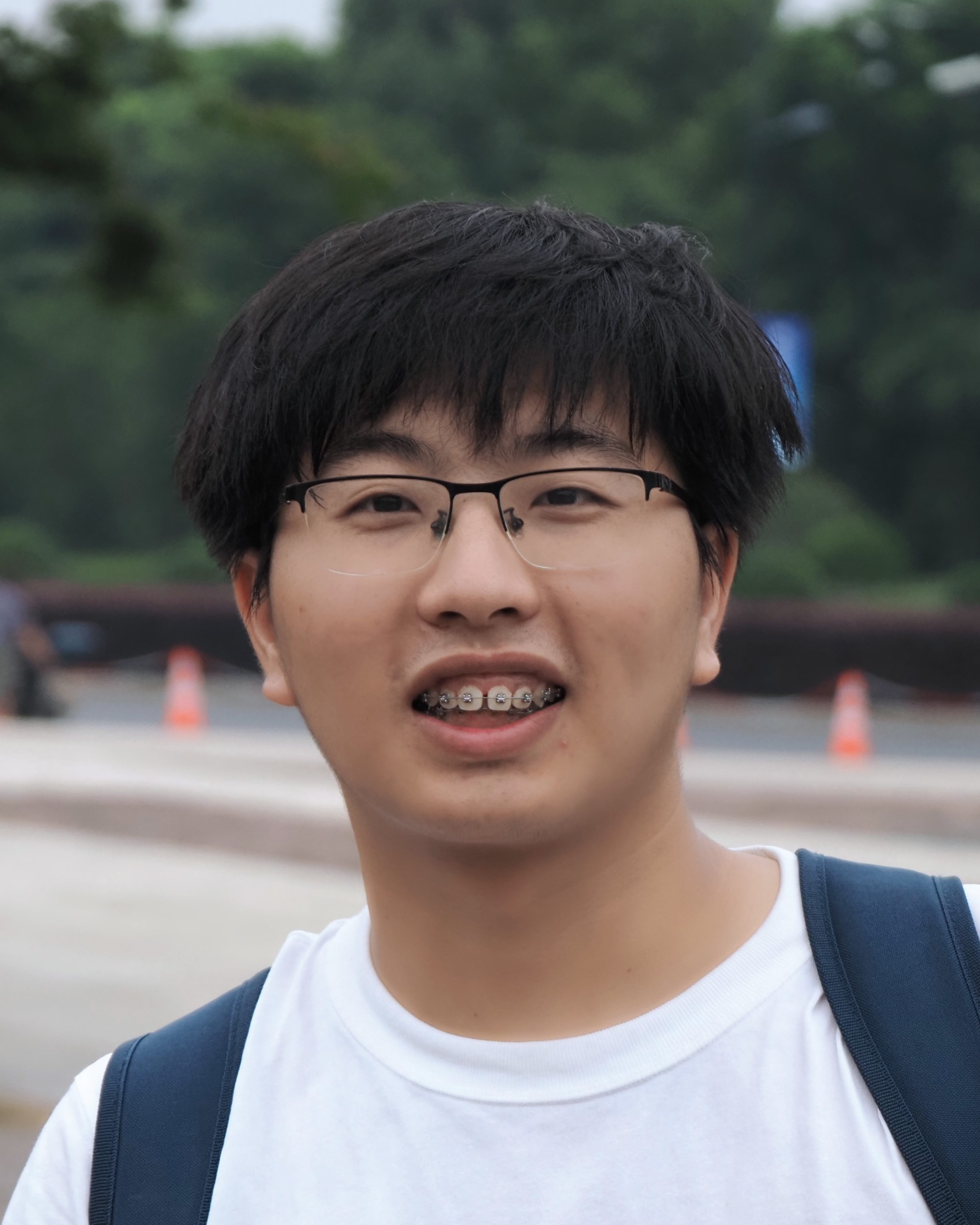}}]{Weicao Deng}
(Student Member, IEEE) recieved the B.Eng. degree from the College of Information Science and Electronic Engineering, Zhejiang University, Hangzhou, China, in 2021.
He is currently pursuing the Ph.D. degree there.
His research interests include millimeter-wave communications, and AI-empowered wireless communication and networking.
\end{IEEEbiography}
\vspace{-10mm}
\begin{IEEEbiography}[{\includegraphics[width=1in,height=1.25in,clip,keepaspectratio]{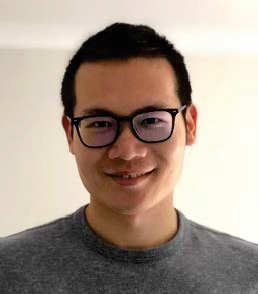}}]{Min Li} 
(Member, IEEE) received the B.E. degree in telecommunications engineering and the M.E. degree in information and communication engineering from Zhejiang University, Hangzhou, China, in June 2006 and June 2008, respectively, and the Ph.D. degree in electrical engineering from Pennsylvania State University, State College, PA, USA, in August 2012. 
He was a Post-Doctoral Fellow with the School of Engineering, Macquarie University, Sydney, Australia, from 2012 to 2016 and from 2018 to 2019, and with the School of Electrical Engineering and Computing, The University of Newcastle, Callaghan, Australia, from 2016 to 2018. 
Since March 2019, he has been a ZJU100 Young Professor with the College of Information Science and Electronic Engineering, Zhejiang University. 
His research interests include network information theory, millimeter-wave cellular communications, integrated sensing and communication systems, and covert wireless communication. 
He has received the Young Rising Star Award by the Information Theory Society of Chinese Institute of Electronics in 2021. 
He was an Exemplary Reviewer of IEEE TRANSACTIONS ON COMMUNICATIONS in 2018 and in 2021. 
He was the TPC Chair for the 18th Australian Communications Theory Workshop (AusCTW) and is the Publicity Chair for 2025 IEEE Information Theory Workshop (ITW).
\end{IEEEbiography}
\vspace{-10mm}
\begin{IEEEbiography}
[{\includegraphics[width=1in,height=1.25in,clip,keepaspectratio]{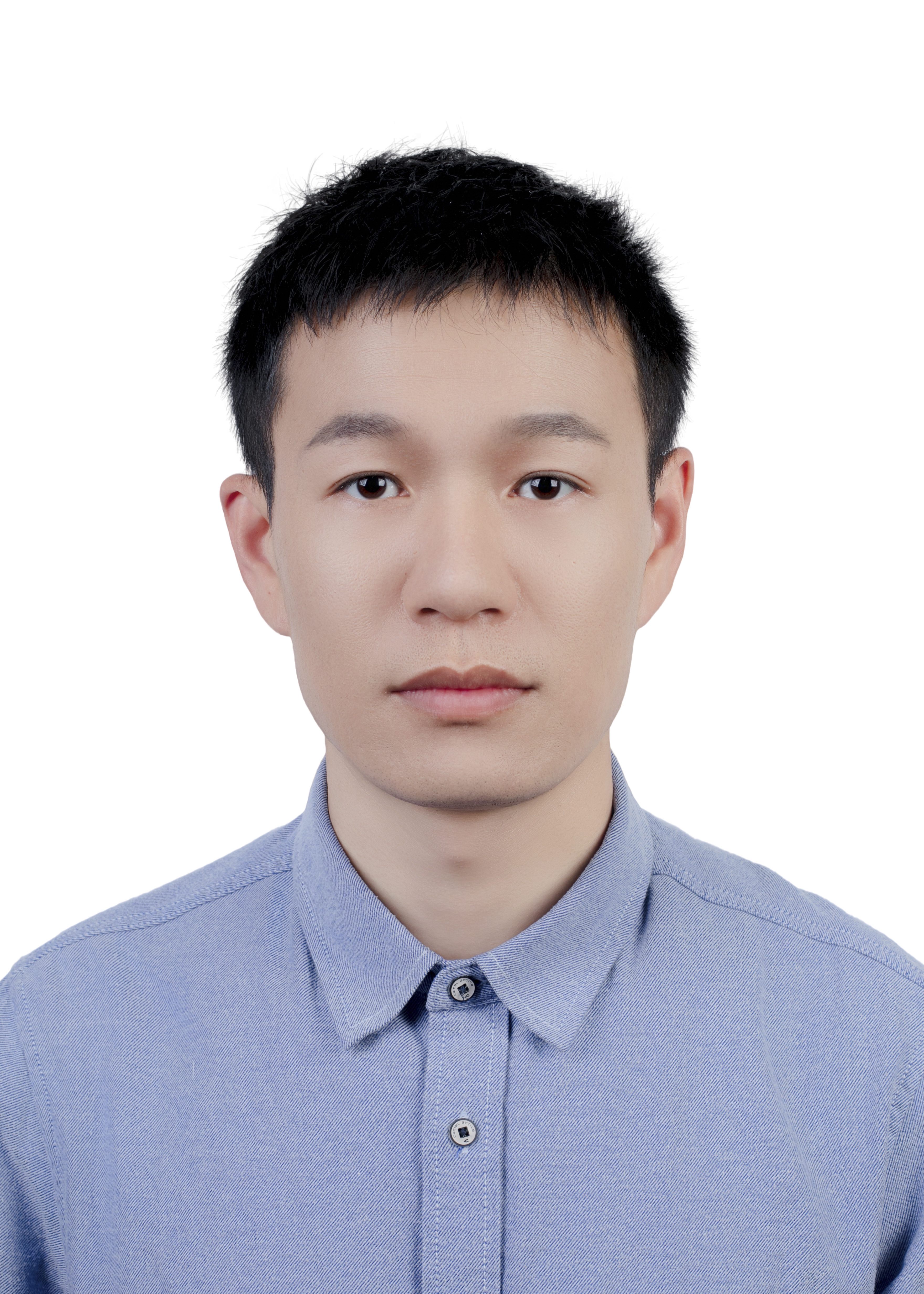}}]{Ming-Min Zhao}
(Senior Member, IEEE) received the B.Eng. and the Ph.D. degrees in Information and Communication Engineering from Zhejiang University in 2012 and 2017, respectively. 
From Dec. 2015 to Aug. 2016, he was a Visiting Scholar at the Department of Electrical and Computer Engineering, Iowa State University, Ames, USA. 
From Jul. 2017 to Jul. 2018, he worked as a Research Engineer at Huawei Technologies Co., Ltd. 
From May 2019 to Jun. 2020, he was a Visiting Scholar at the Department of Electrical and Computer Engineering, National University of Singapore. 
Since Aug. 2018, he has been working with Zhejiang University, where he is currently an Associate Professor with the College of Information Science and Electronic Engineering. 
His research interests include signal processing for communications, channel coding, algorithm design and analysis for advanced MIMO, cooperative communication and machine learning for wireless communications.
He was the recipient of the IEEE Communications Society Katherine Johnson Young Author Best Paper Award in 2024.
\end{IEEEbiography}
\vspace{-10mm}
\begin{IEEEbiography}
[{\includegraphics[width=1in,height=1.25in,clip,keepaspectratio]{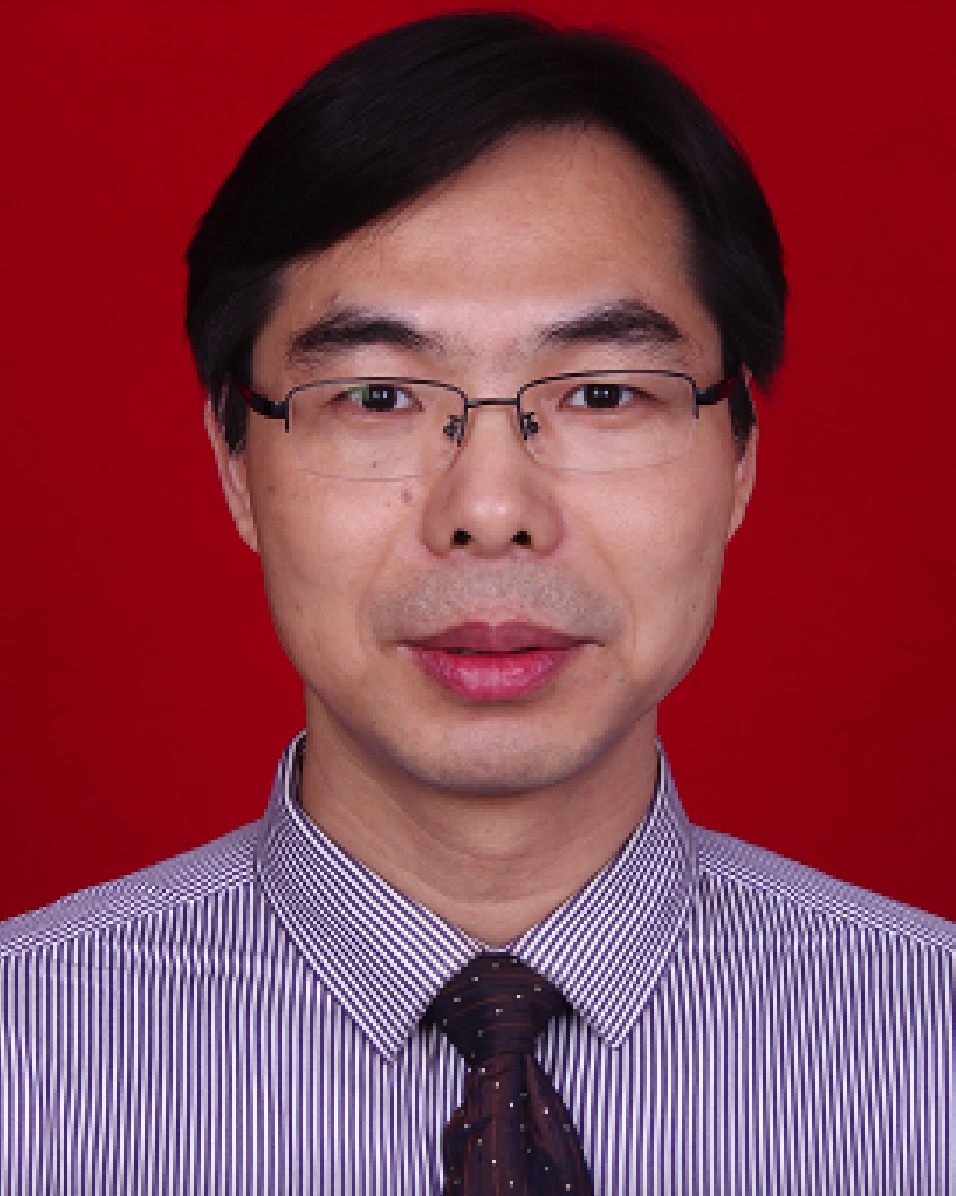}}]{Min-Jian Zhao}
(Member, IEEE) received the M.Sc. and Ph.D. degrees in communication and information systems from Zhejiang University, Hangzhou, China, in 2000 and 2003, respectively. 
He was a Visiting Scholar with the University of York, York, U.K., in 2010. 
He is currently a Professor and the Deputy Director with the College of Information Science and Electronic Engineering, Zhejiang University. His research interests include modulation theory, channel estimation and equalization, MIMO, signal processing for wireless communications, antijamming technology for wireless transmission and networking, and communication SOC chip design.
\end{IEEEbiography}
\vspace{-10mm}
\begin{IEEEbiography}
[{\includegraphics[width=1in,height=1.25in,clip,keepaspectratio]{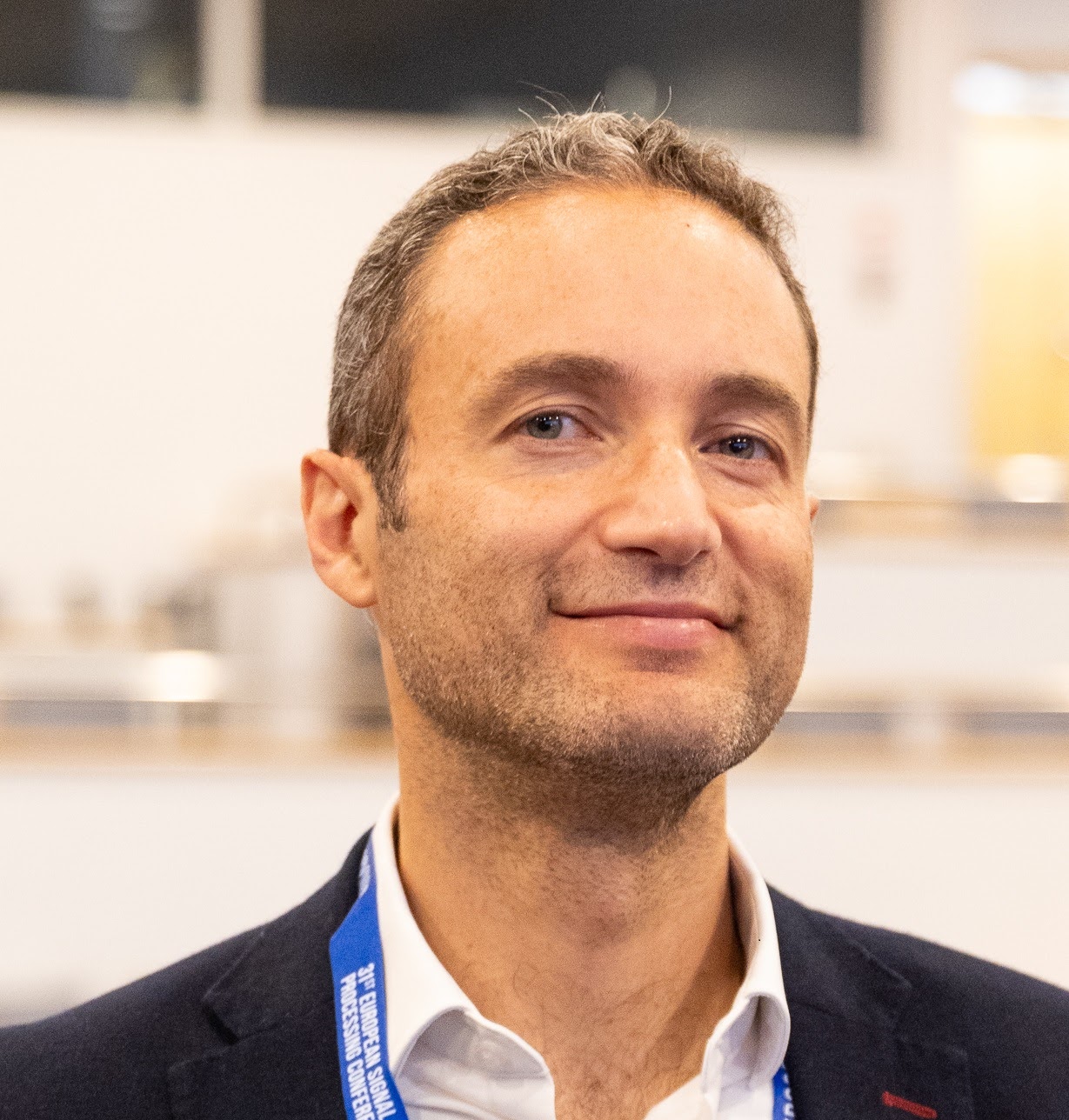}}]{Osvaldo Simeone}
(Fellow, IEEE) is a Professor of Information Engineering. 
He co-directs the Centre for Intelligent Information Processing Systems within the Department of Engineering of King's College London, where he also runs the King's Communications, Learning and Information Processing lab. 
He is also a visiting Professor with the Connectivity Section within the Department of Electronic Systems at Aalborg University. 
He received an M.Sc. degree (with honors) and a Ph.D. degree in information engineering from Politecnico di Milano, Milan, Italy, in 2001 and 2005, respectively. 
From 2006 to 2017, he was a faculty member of the Electrical and Computer Engineering (ECE) Department at New Jersey Institute of Technology (NJIT), where he was affiliated with the Center for Wireless Information Processing (CWiP). 
His research interests include information theory, machine learning, wireless communications, neuromorphic computing, and quantum machine learning. 
Dr Simeone is a co-recipient of the 2022 IEEE Communications Society Outstanding Paper Award, the 2021 IEEE Vehicular Technology Society Jack Neubauer Memorial Award, the 2019 IEEE Communication Society Best Tutorial Paper Award, the 2018 IEEE Signal Processing Best Paper Award, the 2017 JCN Best Paper Award, the 2015 IEEE Communication Society Best Tutorial Paper Award and of the Best Paper Awards of IEEE SPAWC 2007 and IEEE WRECOM 2007. 
He was awarded an Open Fellowship by the EPSRC in 2022 and a Consolidator grant by the European Research Council (ERC) in 2016. 
He is a Fellow of the IET, EPSRC, and IEEE.  
\end{IEEEbiography}

\end{document}